\documentclass[sandspaper,traditabstract,onecolumn]{sands} % for single column
\usepackage{graphicx}
\usepackage{subfigure}
\usepackage{epstopdf}
\usepackage{amsmath,amsfonts,amssymb}
\usepackage[numbers,compress]{natbib}%numbered reference
\usepackage{hyperref}
\usepackage{url}
\usepackage{multirow}
\usepackage{booktabs}
\usepackage{array}
\usepackage{bm}
\usepackage{caption}
\DeclareCaptionFont{scriptsize}{\scriptsize}

\hypersetup{colorlinks=true,citecolor=blue,urlcolor=blue,linkcolor=blue}
\articletype{Review}
\subcateg{Digital Finance} % Information Network / Integrated Circuits / Software Engineering / Industrial Control / Intelligent Transportation / Medical and Healthcare / Digital Finance / Social Governance / Other Fields
\doi{https://doi.org/10.1051/sands/xxxxxxx}
\volume{x}
\pagenos{xxxxxxx}
\newyear{2023}
\citationauthor{Author A, Author B and Author C} % insert in \artcitation{.....}
\begin{document}

\title{Behavioral Authentication for Security and Safety}

%%% if title is too large for page header
\titlerunning{Behavioral Authentication for Security and Safety}

%\subtitle{This is Subtitle of the article}

\author{Cheng Wang\inst{1,2,3*} \and
Hao Tang\inst{1,2} \and Hangyu Zhu\inst{1,2} \and Junhan Zheng\inst{1,2} \and Changjun Jiang\inst{1,2,3}\contact{cjjiang@tongji.edu.cn, cwang@tongji.edu.cn}}

%%% if author is too large for page header
%\contact{cwang@tongji.edu.cn}
\authorrunning{Cheng Wang and Changjun Jiang et al.}

\institute{Department of Computer Science and Technology, Tongji University, Shanghai {\rm 201804}, China \and Key Laboratory of Embedded System and Service Computing, Ministry of Education, Shanghai {\rm 201804}, China \and Shanghai Artificial Intelligence Laboratory, Shanghai {\rm200232}, China.}

\abstract{
The issues of both system security and safety can be dissected integrally from the perspective of behavioral \emph{appropriateness}.
That is, a system is secure or safe can be judged by whether the behavior of certain agent(s) is \emph{appropriate} or not.
Specifically, a so-called \emph{appropriate behavior} involves the right agent performing the right actions at the right time under certain conditions. Then, according to different levels of appropriateness and degrees of custodies, behavioral authentication can be graded into three levels, i.e.,
the authentication of behavioral \emph{Identity}, \emph{Conformity}, and \emph{Benignity}.
In a broad sense, for the security and safety issue,
behavioral authentication is not only an innovative and promising method due to its inherent advantages but also a critical and fundamental problem due to the ubiquity of behavior generation and the necessity of behavior regulation in any system.
By this classification, this review provides a comprehensive examination of the background and  preliminaries of behavioral authentication.
It further summarizes existing research based on their respective focus areas and characteristics. The challenges confronted by current behavioral authentication methods are analyzed, and potential research directions are discussed to promote the diversified and integrated development of behavioral authentication. }

\keywords{behavioral authentication,  security and safety, behavior modeling, anomaly detection, machine learning, artificial intelligence}

\date{Received: xx xxxxx 2022 / Revised: xx xxxxx 2022 / Accepted: xx xxxxx 2022 / Published online: xx xxxxx 2022}

\maketitle

\section{Introduction}\label{Introduction}
%\vspace{-0.1in}
%-------------------------------------------------------------------------------

%\begin{figure}[h]
%\begin{minipage}
%  \includegraphics[width=0.4 \textwidth] {pic/framebass.pdf}
%  \caption{Figure \ref{figure:framebass}(a) illustrates the concept of security can be seen as a ``reasonable state" during the event, a ``reasonable process" (sequence of states) after the event, and a ``reasonable situation" (trend in the development of states) before the event. The essence of studying security lies in examining whether the changing states are secure, and the factors inside and outside the state changes are referred to as behavior.
%  }
%所谓安全，在事中看是一种[合理状态];在事后看一种[合理过程](状态序列) ;在事前看则是一种[合理态势] (状态发展的趋势)研究安全的本质在于研究改变中的状态是否安全，而状态改变的内外要素即为 行为。
%  \label{figure:framebass}
%  \end{minipage}
%%  \vspace{-0.1in} Figure \ref{figure:framebass}(b) describes a technical taxonomy of behavioral authentication.
%\end{figure}    \
\begin{figure}
  \begin{minipage}[b]{0.52\textwidth}
    \includegraphics[width=\textwidth]{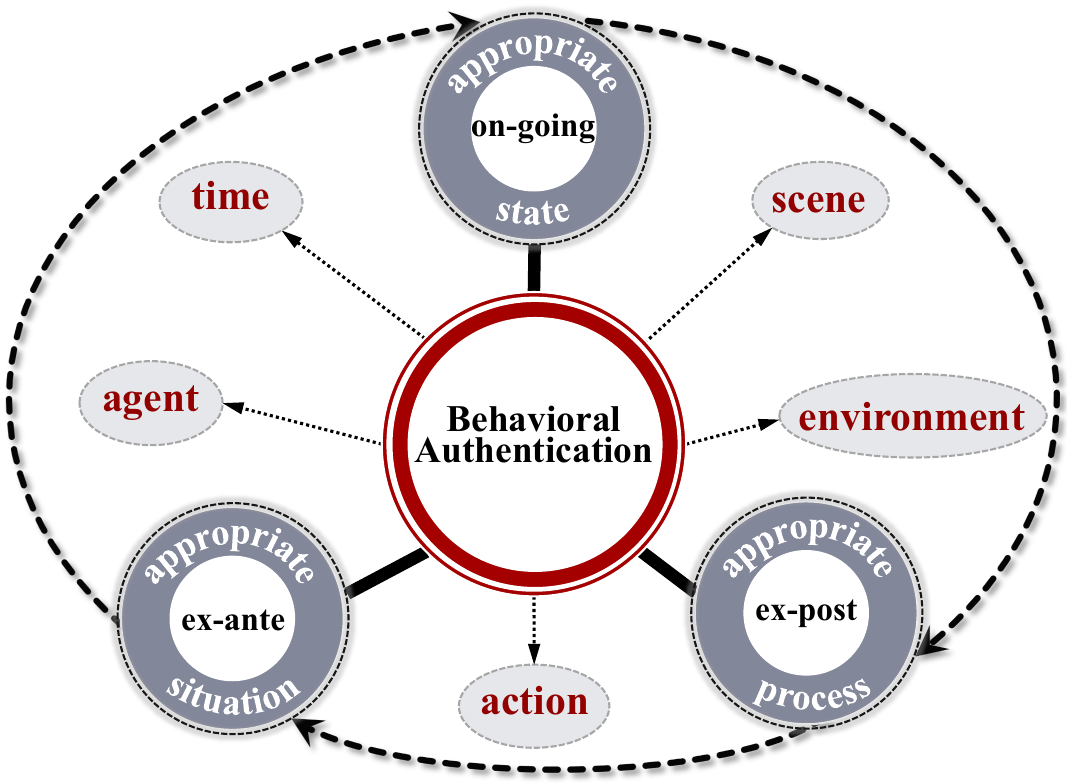}
    \centerline{(a)}
  \end{minipage}%
   \hspace{1cm}
  \begin{minipage}[b]{0.4\textwidth}
    \includegraphics[width=\textwidth]{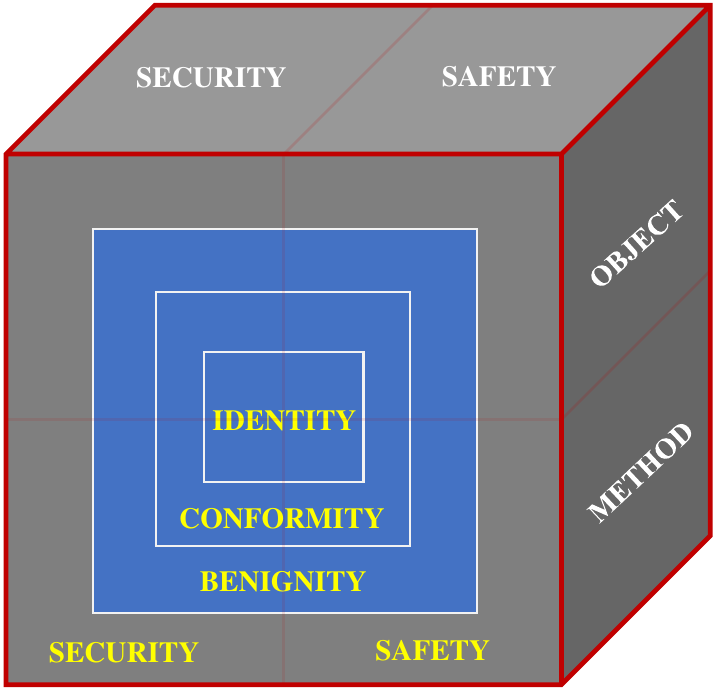}
    \centerline{(b)}
  \end{minipage}
    % \caption{Figure \ref{fig:bass}(a) illustrates the concept of security can be seen as a ``appropriate state" during the event, a ``appropriate process" (sequence of states) after the event, and a ``appropriate situation" (trend in the development of states) before the event. The essence of studying security lies in examining whether the changing states are secure, and the factors inside and outside the state changes are referred to as behavior. Figure \ref{fig:bass}(b) describes the conceptual framework of behavioral authentication.}
     \caption{(a) illustrates the concept of security and safety can be seen as an ``appropriate state" during the on-going detection, an ``appropriate process" (sequence of states) in the ex-post evaluation, and an ``appropriate situation" (trend in the development of states) in the ex-ante awareness. The essence of studying security lies in examining whether the changing states are secure, and the factors inside and outside the state changes are referred to as behavior. (b) describes the conceptual framework of behavioral authentication.}
       \label{fig:bass}
\end{figure}

The advantages of frictionlessness, continuousness, synthesizability and inherence in behavioral authentication have attracted significant attention from security research community \cite{he2022gait2vec,zhang2020deepkey,xu2020touchpass,wang2021framework,wang2021ifdds,kurt2022online,hou2022lightweight,garg2019hybrid,lyastani2018better}. Narrowly defined, behavioral authentication usually refers to \emph{behavioral identity authentication} \cite{tcssWangZY22,IOTJLiYWMJ22}.
It is a technology that has emerged as a result of the continuous development of artificial intelligence industry and  the improvement of computing and storage capabilities in hardware, driven by the increasing dependence of individuals on the convenience provided by devices.
%It is a technology that has emerged as users increasingly rely on the convenience provided by various devices, accompanying the continuous improvement of computing and storage capabilities in terminal devices.
%Unlike traditional knowledge-based identity authentication methods such as PIN codes, picture gestures and passwords \cite{pal2009indian,zhao2013security,bonneau2015passwords,bonneau2012quest}, which are easily lost and stolen,
Behavioral authentication collects behavioral data implicitly in the background and distinguishes them from others by analyzing unique behavior patterns, as behavior patterns are the intrinsic properties of agents and naturally possess distinct characteristics compared to others.
Initially, research on behavioral identity authentication primarily focused on utilizing collected behavioral data to address specific authentication scenarios and meet the essential usability requirements \cite{yang2017energy,shi2022identity,wu2021network}.
%Initially, research on behavior identity authentication mainly treats it as a method and applies the collected behavioral data to corresponding authentication scenarios \cite{yang2017energy,shi2022identity,wu2021network}.
As researchers gain a deeper understanding of behavior, they realize the close correlation between security and behavior \cite{perera2018face,shen2018performance}. The connotation of security can be described from \emph{a behavioral perspective},
where it involves the \emph{right agent} performing the \emph{right actions} at the \emph{right time} under certain conditions (e.g., \emph{scene} and \emph{environment}), as illustrated in Figure \ref{fig:bass}(a).
Therefore, the study on security can be essentially implemented by the research on behavior, then the behavior is also regarded as the object of security research \cite{lee2019parameterized,zou2020deep}.
Behavioral identity authentication ensures protection against external malicious attackers and safeguards security.
As the scope of  behavioral identity authentication applications gradually expands, the importance of non-functional requirements is also increasing. Researchers are paying more attention to the robustness, reliability, and timeliness aspects of behavioral identity authentication \cite{zhang2020deepkey,zou2020deep,yang2019behavesense}, which is of great significance to the safety.

%Some research also focuses on ensuring the robustness and usability of behavioral identity authentication schemes themselves \cite{zou2020deep,yang2019behavesense,zhang2020deepkey},
%Therefore, behavioral authentication is inherently intertwined with system safety and security.

%However, with the increasing requirements for authentication accuracy, some studies have treated behavior as an object, exploring how to synthesize more reliable behavior representations for user verification . The trend of integrating behavior as both a method and an object becomes more evident, enabling more precise behavior characterization and customization for different scenarios.
%

%但是我们在实际的业务过程当中，发现有一些针对行为的问题不属于行为身份认证。例如在借贷反欺诈场景中，有一些用户使用的是自己真实的身份，但是在通过后，他们会进行恶意套现。 我们称这类问题的研究为

In real-life business, after verifying the legitimacy of identity, there are still some behaviors that require further behavioral authentication to  evaluate their compliance.
For example, in the scenario of credit loans \cite{WangZHLJ2343242}, there are some users who use their genuine identities, but after being approved, they engage in malicious cash-out activities.  In the database system, users access the system through their own account, so the user's identity is normal, but when the user steals database information through this account, the behavior is illegal for the system.
 We call this type of problems as \emph{behavioral conformity authentication}. Behavioral conformity authentication refers to identifying whether behavior patterns conform to the rules within the system during the process of using various intelligent information services under the precondition of user legitimate identity. In fact, some studies have already utilized behavioral data as the resource to solve different cybersecurity issues, e.g., online payment anti-fraud \cite{WangZHLJ2343242,wang2022caesar,tdscWangWZC2123423}, network system intrusion detection \cite{wang2022wrongdoing}, and social network compromised account detection \cite{tcssWangZY22,ring2019survey,shu2020fakenewsnet}.
%With the further development of 5G and Internet of Things (IoT) technologies, real-time behavioral data is generated and recorded across different systems. The definition of behavioral authentication is no longer limited to behavior identity authentication.The behavioral authentication proposed in this paper refers to a broad concept of behavioral authentication, encompassing not only behavior identity authentication but also compliance authentication and trust authentication.

Furthermore, security and safety issues may arise due to unknown risks and vulnerabilities associated with an agent's behavior, even if its identity is legitimate and its actions comply with regulations.
%Furthermore, certain security and safety concerns exist because of unknown risks and vulnerabilities associated with an agent's behavior, even if its identity is legitimate and its actions are in compliance with regulations.
%Furthermore, certain security and safety concerns exist where the agent's identity is legitimate and its actions are in compliance with regulations, but there may be unknown risks and vulnerabilities associated with its behavior.
For instance, in the scenario of credit loans, there is the issue of multiple loans, where some users borrow from different platforms using their genuine identities. Their identities are legitimate, and they repay the loans on time initially.
However, due to the limited repayment capacity of some borrowers, there is a higher risk associated with borrowing from multiple platforms. In the short term, it may seem like there are no issues, but when the last platform is unable to borrow, it leads to overdue payments and a mounting debt burden, resulting in bankruptcy caused by a cycle of borrowing to repay existing debts.
In an industrial Internet scenario, the operational state of a device manifests as an uninterrupted flow of temporal data, and occasionally accompanied by abnormal signals do not trigger compliance alerts. These signals do not originate from external intrusions upon the device,  but the accumulation of occasional abnormal signals over the long term can lead to systemic risks of system \cite{charles2017loan,bailey2022fraudulent,tdscsdfWangZ22}. These concerns are called  \emph{behavioral benignity authentication}.
It refers to detecting any potential risks in user behavior during the process of using various intelligent information services under the precondition of user identity legitimacy and behavioral compliance.

%Considering the aforementioned aspects, behavioral authentication consists of behavior identity authentication, behavior conformity authentication, and behavior benignity authentication, which is shown in Figure \ref{figure:framebass}.
%In this work, we first provide an overview of the background of behavioral authentication and conclude the limitations of existing authentication methods.
%Next, we summarize the existing research and analyze the characteristics of relevant studies within each category.
%Additionally, we identify the limitations of current behavioral authentication methods and distill future research directions.

Considering the aforementioned aspects, behavioral authentication encompasses three main components: behavioral identity authentication, behavioral conformity authentication, and behavioral benignity authentication. We propose the conceptual framework of behavioral authentication, which is shown in Figure \ref{fig:bass}(b).
In this work, we first provide an overview of the background and  preliminaries of behavioral authentication.
Next, we conduct a comprehensive review of existing research and categorize the studies based on their respective focus areas and distinctive characteristics within each category.
Moreover, we highlight and discuss the challenges of the current behavioral authentication methods,
 and propose future research directions to enhance the effectiveness and efficiency of behavioral authentication.
%Considering the aforementioned aspects, including behavior identity authentication, behavior conformity authentication, and behavior benignity authentication, we propose the integrated behavioral authentication framework, which is shown in Figure \ref{figure:framebass}.
%In this paper

%This isolated and heterogeneous approach is prone to create loopholes and backdoor attacks.
%The proposed integrated approach to behavioral authentication in order to enhance safety and security of the system.

%

\section{Background and preliminaries}\label{Background}

%\vspace{-0.1in}
%-------------------------------------------------------------------------------
%Cybercriminals have easy access to, steal, or purchase personal data such as email and physical addresses, phone numbers, birthdates, and other personally identifiable information, allowing them to gain unauthorized access or create fraudulent accounts.
%Malicious software, remote access tools, and other technologies employed by cybercriminals exploit vulnerabilities in passwords, device IDs, one-time passwords, and other authentication mechanisms.
%Overall, the number and audacity of cybercriminals are on the rise.
%    Each passing day brings forth emerging threats, rendering traditional security practices ineffective. The need of the hour is to adopt novel technologies to counter these attacks. Among the most promising advancements is the field of behavioral authentication, which not only revolutionizes user authentication but also prioritizes privacy. This groundbreaking approach is reshaping the security industry.
\subsection{Background}
According to the $52th$ \emph{Statistical Report on the Development of the Internet in China} \cite{2023china} released by the China Internet Network Information Center in $2023$, as of June $2023$,  China's internet user base has reached $1.079$ billion people, with an internet penetration rate of $76.4\%$. The user base for instant messaging, online video, and short video stands at $1.047$ billion, $1.044$ billion, and $1.026$ billion, respectively, with user adoption rates of $97.1\%$, $96.8\%$, and $95.2\%$. These new digital services have been closely related to people's lives.
%the user base of online communication services and online video services reached $1.007$ billion and $975$ million respectively. Emerging types of internet services have also shown a rapid growth trend, such as remote office and connected car services with a year-on-year growth rate of $35.7\%$ and $23.9\%$ respectively. These new digital services have become closely intertwined with people's lives.
%我国网民规模达到了10.79亿人，互联网普及率达76.4%。即时通信、网络视频、短视频用户规模分别达10.47亿人、10.44亿人和10.26亿人，用户使用率分别为97.1%、96.8%和95.2%。
However, at the same time, digital transformation is also facing severe and complex security and safety challenges, posing real threats to critical infrastructure, systems, and citizen privacy \cite{ChenWYHJ23,ChenWCYHJ2345}. From the perspective of attack targets, the focus has shifted from networks and systems to business and data, with an increasing number of ransomware and application-based attacks. In terms of attack methods, the trends of automation, intelligence, and concealment in attack tools have become more prominent. Additionally, cybercriminals further enhance the success rate of attacks by illegally acquiring personal data, such as email and physical addresses, phone numbers, and other personally identifiable information. All these factors pose significant challenges to traditional authentication methods. Driven by the security and safety demands of the real world, there is a current need to adopt new technologies to address these challenges. Behavioral authentication is one of the most promising methods, and this innovative approach is moving towards a safer and more secure cyberspace.

The technique of behavior modeling is closely related to behavioral authentication.
%In fact, research on predicting user characteristics (attributes, personality, and behavior) through user behavior modeling has a long history. In $2013$ and $2015$, Kosinski et al. \cite{youyou2015computer,kosinski2013private}  explored the feasibility of using user behavioral data to infer, predict, and model user attributes, interests, and personality.
In fact, there is a rich history of research focused on predicting user characteristics, including attributes, personality, and behavior, through the utilization of behavior modeling  techniques. In the years $2013$ and $2015$, Kosinski et al. \cite{youyou2015computer,kosinski2013private} explored the feasibility of using user behavioral data to infer, predict, and model user attributes, interests, and personality.
 In terms of user behavior prediction, as early as $2010$, Song et al. \cite{song2010limits} conducted a three-month study and analysis of the travel records of $50,000$ anonymous mobile phone users in their publication in \emph{Science}, and found that users' historical travel behaviors followed specific patterns, with an accuracy of up to $93\%$ in predicting users' potential travel behavior.
There has been extensive research in the specific application of behavior modeling requirements for internet services. For example, personalized modeling has been implemented by mining users' interests and hobbies based on their behavioral characteristics \cite{yin2016discovering,hashemi2017go,zhou2018atrank,wang2018tsaub}. Behavior trend analysis has been conducted by observing changes in user behavior patterns \cite{nai2018modeling,yang2014characterizing,huo2017user,zhu2017next}. Malicious accounts have been detected by analyzing account behavioral characteristics \cite{cao2014uncovering,zhang2015research}. Risk assessment of default has been performed by analyzing user transaction behavior records \cite{wang2018novel}. Social identity association across different platforms has been addressed by matching user behavior patterns \cite{liu2015structured,wang2018comprehensive,zafarani2013connecting}.

In the field of identity authentication, the majority of current online methods resemble access control methods. Common methods include setting up alphanumeric or graphical passwords for accounts \cite{pearman2017let,ye2017cracking,constantinides2020eye,constantinides2019accuracy,pal2009indian,bonneau2015passwords}, utilizing security tokens \cite{turner2008implementing,hallsteinsen2007using,pal2009indian,bonneau2012quest}, and employing biometric features such as facial and iris recognition \cite{ruiz2016cerebre,sitova2015hmog,bicego2006use,clancy2003secure,kumar2010comparison}. Setting passwords for accounts is the simplest and most widely used method of identity authentication, while linking mobile phones and emails  serves as an effective measure for account protection and recovery. Security tokens are physical devices employed as a means of identity authentication and are currently extensively utilized. Biometric authentication refers to the utilization of unique biological features of individuals to verify their identities. Biometric authentication is considered a relatively reliable method of identity verification.
The above-mentioned methods primarily operate during the login authentication phase of an account. Due to the fact that these methods typically require additional user operations, they are regarded as intrusive authentication methods. These intrusive methods are hard to meet the high requirements of users for authentication convenience and service experience.
%The aforementioned methods primarily operate during the login authentication phase of an account. As these methods typically require user operations, they are considered intrusive protective methods. However, to ensure user experience, these intrusive protective methods often fail to meet the requirements of continuous authentication.
When confronted with the issue of managing multiple account passwords, many users tend to adopt complete or partial password reuse strategies, underestimating the threat of password reuse to account security \cite{das2014tangled,sun2011opass,lyastani2018better}. Furthermore, certain risks exist with certain biometric-based authentication technologies. For instance, in facial recognition, the replicability of facial data (obtainable and replicable in public environments), the instability of facial data (affected by makeup, allergies, injuries, and cosmetic surgeries leading to changes in features), and the security of backend data should be taken seriously (a breach of backend data would have devastating consequences for industries and society).

In the field of conformity authentication, the existing practices mainly rely on establishing conformity authentication systems based on rules \cite{xu2005fuzzy,cao2010domain,tandon2003learning,pan2015anomaly,li2007roam}. The process of setting up rule libraries is time-consuming and incurs high manual costs \cite{lin2020anomaly}. Once authentication is granted through rule-based detection systems, these systems tend to persist in granting access for similar requests in the future.  However, due to the absence of timely updates to the rule library, identifying new instances of non-conformity becomes challenging.
%Generally, once authentication is granted through such rule-based detection systems, the system will continue to grant access for similar requests in the future. Due to the lack of timely updates to the rule library, it becomes difficult to identify new non-conformities.
In practical business scenarios, the limitations of rule-based construction have become increasingly apparent. Conformity authentication models sometimes require the input of hundreds of expert features. The manual construction of rules is also challenging to transfer and switch across different application scenarios, resulting in significant manual and time costs and impacting the efficiency of model development and operation. This method is constrained by human expertise and may overlook potential non-conformities, preventing the model from achieving excellent detection performance \cite{tandon2007weighting}. Additionally, this time-consuming and labor-intensive approach to constructing conformity models, which may overlook complex risks, no longer meets the requirements of secure, dynamically changing conformity authentication systems. It is of great significance for the development of conformity authentication models to effectively utilize and reuse knowledge, reduce manual and time costs, and establish automated and efficient authentication models.

In terms of benignity authentication, existing methods mainly focus on
the static benignity of entities, with limited scalability in dynamic environments. These methods heavily rely on pre-established dependency analysis, equivalence relationships, and protocol state specifications.
The trusted formal modeling employed in these methods is predominantly based on cryptographic techniques, which presents several challenges in real-world business scenarios. Cryptographic modeling processes can be time-consuming and hinder the efficiency of authentication procedures.
The practicality of these cryptographic methods may be limited due to high hardware requirements for deployment, affecting their efficiency and usability.
These methods lack the necessary adaptability and flexibility, which makes it challenging to promptly capture potential risks. For example, coupon clippers may not initially violate regulations, but the large-scale organization of coupon clippers' activities is likely to increase overall transaction and credit costs,  ultimately hindering normal business activities.
%Consequently, coupon clipping is deemed non-benign behavior.
There is a pressing need for more adaptive and flexible approaches in benignity authentication that can overcome these limitations and meet the demands of dynamic environments in real-world business scenarios.
%In terms of benignity authentication, existing methods mainly focus on the static benignity of individual entities, and determining the benignity often requires a trusted third party to execute. These methods heavily rely on pre-established dependency analysis, equivalence relationships, and protocol state specifications.
%The trusted formal modeling employed in these methods is predominantly based on cryptographic techniques, which poses several challenges in real-world business scenarios.  The process of cryptographic modeling is time-consuming and can impede the efficiency of authentication procedures.  The practicality of these cryptographic approaches may be limited in real-world contexts, where efficiency and usability are crucial considerations.
%These methods lack adaptability and flexibility, which makes it difficult to effectively capture potential risks in a timely manner. For example, taking advantage of loopholes may not violate the regulations initially, but in the long run, it poses risks to the system. In reality, it is considered as non-benignity behavior.
%There is a need for more adaptive and flexible approaches in benignity authentication that can address these limitations and meet the requirements of dynamic environments in real-world business scenarios.
%这些方法缺乏适应性和灵活性难以即使捕获潜在的风险

\subsection{Preliminaries}
Experts in the field of cybersecurity have  paid attention to the importance of user behavior research in addressing security issues. Dr. Douglas Maughan, the inaugural Office Head for the National Science Foundation (NSF) Convergence Accelerator, and the former Director of the Cyber Security Division at the Science and Technology Directorate of the U.S. Department of Homeland Security, believes that researchers should prioritize viewing cybersecurity issues from a human factors perspective \cite{waldrop2016cybercrime}. Professor Angela Sasse of University College London, funded by the British government, is studying cybersecurity issues in the business sector from a novel perspective of social and behavioral science \cite{krol2016towards}. Professor Stefan Savage of the University of California, San Diego is also conducting research on behavior analysis for the prevention and control of network fraud \cite{pearce2014characterizing}. \emph{Nature} published an article titled \emph{How to Hack The Hackers: The Human Side of Cybercrime}, highlighting the progress made in representative research efforts and emphasizing the importance of leveraging behavioral science to understand the behavior patterns of both perpetrators and victims for enhancing cybersecurity \cite{waldrop2016hack}.

For behavioral identity authentication, devices typically collect behavioral data in the background.
The collected data is then utilized for training a machine learning model \cite{abuhamad2020sensor}.  Features are extracted through the trained machine learning model based on the collected behavioral data, which form the users' behavior profiles.
The identity behavior profile is represented as a matrix $U$ by the model.
%\begin{equation}
%U=[u_{ij}]_{m\times n}=
%\begin{bmatrix}u_{11}&u_{12}&\cdots&u_{1n}\\
%u_{21}&u_{22}&\cdots&u_{2n}\\
%\vdots&\vdots&\vdots\\
%u_{m1}&u_{m2}&\cdots&u_{mn}
%\end{bmatrix},
%\end{equation}
%where the column $j$ of the matrix represents identity embedding vector $j$, $j= 1, 2, \cdots, n$; the row $i$ of the matrix represents the $i$-dimension feature of identity embedding vectors, $i= 1, 2, \cdots, m$; the element $u_{ij}$ of the matrix represents the $i$-dimension feature of identity embedding vector $u_j$.
During the process of behavioral identity authentication, the newly generated credential of $k$th user is compared against stored identity behavior profile matrix to determine the authenticity of the user's identity.  Verification involves comparing a provided behavior record with a stored template to determine a level of similarity. It  grants access to legitimate users if their presented behavior record exhibits a similarity measure surpassing a predefined threshold. Let $x$, $g(\cdot)$ and $d_k$ denote the behavior record, the behavioral identity authentication model and the predefined threshold of user $k$, respectively.
The result of behavioral identity authentication model $R=True$ if $g(x)>d_k$ and $False$ otherwise.
%查一下行为身份认证的定义
%对于行为身份认证而言，设备通常会在后台进行行为数据的收集，然后基于这些手机的数据训练一个机器学习模型，然后基于训练好的机器学习模型，在对后台收集的行为数据进行特征的提取，从而形成用户的行为模板，在进行行为身份认证的过程中，会将新生成的身份凭证与存储的身份信息进行核对，判断用户的身份是否真实。
%\subsection{The definition of behavioral conformity authentication}
%对于行为合规认证而言，需基于所收集的行为数据，区分用户的身份是否合法，这里参考一下异常检测中的形式化定义。如图2所示，这里的用户身份是合法的，但是他的行为是不守规的，因为他进行了恶意的套现。画一些图来说明下。
%
%
%\subsection{The definition of behavioral benignity authentication}
%对于行为良性认证而言，需基于xxx，如图3所示，一个多头借贷的例子。

For behavioral conformity authentication, it refers to the process of ensuring that the internal behavior of users and business activities comply with relevant laws, regulations, and industry standards.
 Let $\mathcal{X}=\{x_1,x_2,\cdots,x_N\}$ denote the set of behavior records. Behavioral conformity authentication aims to learn a decision function $\phi(\cdot)$: $\mathcal{X}\rightarrow \mathbb{R}$ that assigns behavioral conformity scores. The goal is to effectively distinguish non-compliant behavior records from compliant behavior records in the space defined by the behavioral conformity score decision function. By inputting behavioral data into $\phi(\cdot)$, it can directly infer the conformity scores. Larger outputs of $\phi(\cdot)$ indicate a greater degree of non-compliance,  which requires the system to take appropriate actions timely to prevent or rectify such behaviors. Certainly, it is necessary to characterize and map the original features of behavior records before inputting them,  which is beneficial for obtaining more accurate scores of behavioral compliance.

%Current research in behavioral benignity authentication 的表达，仅仅是有研究者探索到了 良性范围内的安全问题
%Behavioral benignity authentication has not yet formed a clear definition.

Researchers have already begun exploring security and safety issues within the scope of behavioral  benignity authentication.
 Relevant research works primarily focus on the following aspects, including traceability of behavior, predictability
of risk, consistency of execution, and integrity of record. The traceability of behavior refers to the tracking  of individual or entity behavior activities to obtain detailed historical behavioral data, which is helpful for verifying the legitimacy of user behavior. Predictability
of risk involves a comprehensive evaluation of individual or entity behavior activities to timely identify potentially risky behaviors and update the scope of compliance.
Consistency of execution  evaluates credibility and detects potential risks without affecting consistent operations in cross-domain  interactions over heterogeneous networks.
Integrity of record entails protecting user privacy and ensuring consistency in the certification process.
These aspects of behavioral benignity authentication provide additional protection for systems.

%行为认证的不同级别保障金融贷款业务的safety和security的一个典型案例。

%This process involves obtaining certification from third-party or professional certification bodies for their behavior records. Let $\mathcal{X}=\{x_1,x_2,\cdots,x_N\}$ denote the behavior record set. Then, behavioral conformity authentication aims at learning behavioral conformity score decision function $\phi(\cdot)$: $\mathcal{X}\rightarrow \mathbb{R}$ in a way that non-compliant behavior records can be easily differentiated from the compliant behavior records in the space yielded by the behavioral conformity score decision function. $\phi(\cdot)$
%can directly infer the conformity scores with behavioral data inputs. Larger
%outputs of $\phi(\cdot)$ indicate greater degree of being non-compliant.

%行为认证方法从其技术理念上来看主要具有如下优势：1）该方法是一个完全后台的程序，所以是非侵入式的；其收集数据和验证的过程均不需要用户操作。2）该方法将安全认证由一次性变为持续性的过程。3）该方法直接捕获的是行为多个子空间的投影，行为子空间之间具有可合成性;
%4）该方法难以被窃取，因为入侵者几乎不可能非常完美的模仿主人的行为习惯；而且，入侵者盗取利益的侵害行为通常就与用户正常行为相异。 continuousness, synthesizability and inherence

The behavioral authentication method has the following advantages based on its technical principles:
\begin{itemize}
\item \textbf{Frictionlessness.} This method operates as a backend program, eliminating the need for user intervention. The collection of data and verification are seamlessly performed without requiring any explicit actions from the user. This frictionlessness of behavioral authentication ensures a smooth and convenient experience.
\item \textbf{Continuousness.} The method transforms cybersecurity authentication from a one-time process to a continuous one.    It allows for ongoing analysis of user behavior over time, ensuring that access remains granted as long as the behavior patterns align with the established user profile.  This continuous authentication approach enhances security by detecting any anomalous or suspicious behavior in real-time.
\item \textbf{Synthesizability.} The method directly captures the projection of multiple subspaces of behavior, which exhibits synthesizability among the subspaces. By capturing various dimensions of behavior, it creates a comprehensive and reliable behavior profile.  Synthesizability of behavior enhances the accuracy and robustness of the authentication process.
\item \textbf{Inherence.} The data used for behavioral authentication typically originates from users' inherent characteristics and patterns.  Therefore, intruders are unlikely to perfectly mimic the user's behavior patterns, as individual behavior is unique and difficult to imitate convincingly.  Furthermore, intruders typically exhibit intrusive actions aimed at stealing benefits, which often deviate from the user's normal behavior.
\end{itemize}
\begin{figure}[t]
 % \centering
  \includegraphics[width=1 \textwidth] {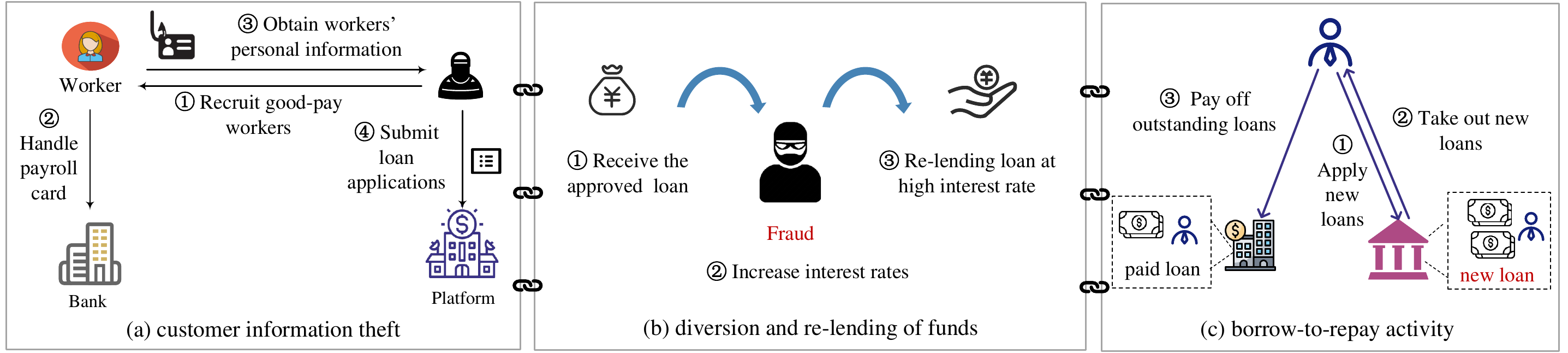}
% \vspace{-0.18in}
\caption{A typical example demonstrating how different levels of behavioral authentication ensure the safety and security of credit loan services. (a) shows criminals collect information from victims and submit loan applications, and it can be addressed by behavioral identity authentication. (b) describes the illicit re-lending of funds activities that occur after the loan has been approved and disbursed, which belongs to the level of behavioral conformity authentication.
In the auditing process, timely analyzing the flow of loans among different platforms and ensuring the predictability of lending risks, are the key issues addressed by behavioral benignity authentication, as shown in (c).
Behavioral authentication expands through a progressive framework of sub-functions, i.e., identity, conformity, and benignity, to achieve gradual enhancement for the safety and security of credit loan services.}
  \label{figure:exa1}
%  \vspace{-0.1in}
\end{figure}

We consider typical scenarios to illustrate the connections between different authentication methods. For example, credit loan services are typical data-intensive services, and the synthetic and intrinsic characteristics of behavioral authentication provide support for enhancing representational capabilities and breaking down isolated data attributes of credit loan services.
During the initial stages of loan application, some criminals may steal others' information through illegal means for fraudulent loan applications. The specific process is illustrated in Figure \ref{figure:exa1}(a). The evaluation of whether the information submitted by a borrower matches the true circumstances falls under the level of behavioral identity authentication, which protects the security of the loan application for financial institutions. While behavioral identity authentication can ensure the legitimacy of a borrower's identity, there are instances of non-compliant behavior after the loan is approved. Diversion and re-lending of funds is one such non-compliant behavior where the borrower does not use the loan funds as intended during the loan application. Instead, after receiving the loan from a financial institution, the borrower engages in illegal lending activities. As shown in Figure \ref{figure:exa1}(b), the fraudster further increases the loan interest rates approved by financial institutions and lends the funds to others to profit from the interest rate differential. Detecting such frauds falls under the level of behavioral conformity authentication. It uses methods like behavior tracking to ensure the security of loan flows. Furthermore, in real business processes, there are cases where a borrower's identity is legitimate, and the borrower's behavior complies with platform regulations, but failure to restrict such behavior in a timely manner can increase overall system risks.  Figure \ref{figure:exa1}(c) illustrates the concept of borrow-to-repay  activity, which is detrimental to the stable operation of the financial platform. Initially, the borrower only applies for loans on a single platform and can adhere to the repayment policies of that platform. However, later on, due to financial difficulties, the borrower cannot repay the loans on one platform  on time. As a result, the borrower applies for new loans on other platforms to repay the outstanding loans. In reality, the borrower's ability to repay loans has declined. The collaborative audit across multiple platforms to promptly detect such behavior not only ensures the security of the business but also further ensures platform safety by timely refining loan strategies and making the controllability of platform risks. They are crucial aspects of behavioral benignity authentication.
Safeguarding user identities from theft, monitoring the compliant use of loans, and continuously improving the approval processes provide  essential protection for credit loan services.
The progressive enhancement of financial market security and safety through the expansion of behavioral authentication sub-functions with the level of identity, conformity, and benignity promotes the development of inclusive finance.

\begin{figure}[t]
  \centering
  \includegraphics[width=1 \textwidth] {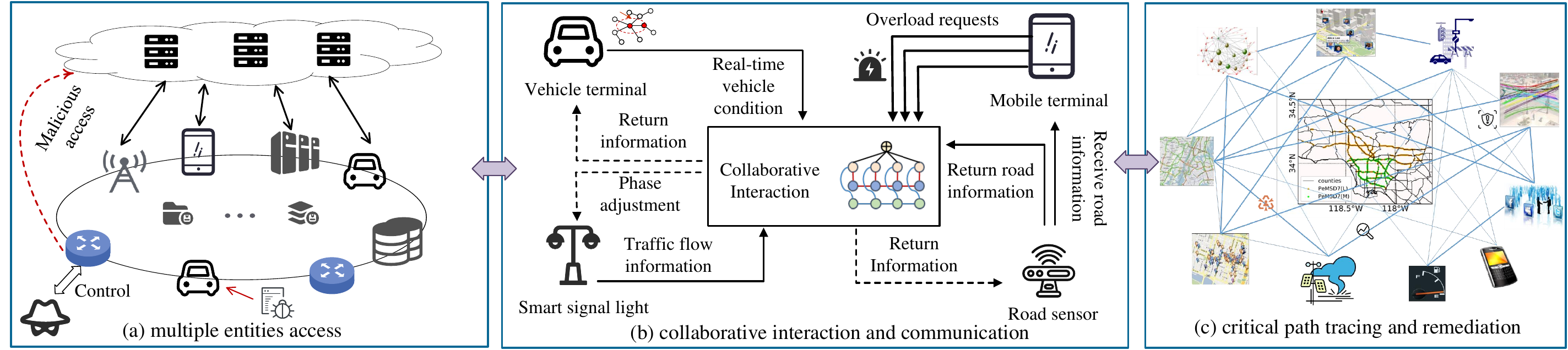}
% \vspace{-0.18in}
  \caption{An example illustrating the security and safety issues that different levels of behavioral authentication need to confront in intelligent transportation information services. (a) describes  the entity access phase where various entities may face threats such as identity theft and phishing attacks, and it belongs to the level of behavioral identity authentication. (b) illustrates the monitoring of API interface traffic  and blocking of non-compliant request entities during collaborative interactions and communication process, which falls under the level of behavioral conformity authentication. (c) shows the functionality of behavioral benignity authentication, which identifies and rectifies high-risk paths within the system, enabling internal information protection  and proactive defense against malicious activities.
   }
  \label{figure:exa2}
\end{figure}

Intelligent transportation information services also involve the extension of different levels of behavioral authentication. As  communication and computation-intensive services, the frictionless and continuous characteristics of behavioral authentication are helpful in addressing emerging security and safety challenges, where  entities' access is dynamic, devices interact frequently, and cross-domain paths are concealed. During the access phase of different entities,  behavioral identity authentication prevents different devices from falling victim to phishing attacks and identity theft, ensuring data and access security for intelligent devices, which is shown in Figure \ref{figure:exa2}(a). While some entities have successfully verified their legal identities and registered successfully within the intelligent transportation system, during the process of heterogeneous entity interaction,  some entities may not adhere to the standards and regulations of services. Figure \ref{figure:exa2}(b) illustrates typical non-compliant behavior in intelligent transportation information services. During the collaborative development of intelligent transportation information system, for the convenience of information sharing and resource coordination, different entities open certain interfaces to other interacting entities, based on additional standards. These entities must conform to the specified usage of external cooperative entities' API interfaces according to these additional standards to ensure efficient and stable information sharing and resource allocation in the system.  There are some access entities that have verified the legitimacy of their identity but generate traffic requests that exceed service standards.
This directly leads to network congestion and increases the system's response time. %Monitoring of interaction traffic falls within the scope of behavioral conformity authentication. Nevertheless, there are some access entities that have verified the legitimacy of their identity and generate traffic requests that exceed service standards. This directly leads to congestion at the system level.
 Behavioral conformity authentication is responsible for monitoring the traffic of API interfaces, promptly detecting non-compliant request entities, and safeguarding the overall security and stability of the multi-entity collaborative process. Additionally, some entities have legitimate identities and generate traffic that complies with standards when using open API interfaces. However, in the process of cross-domain access, there may exist potential high-risk access paths. These paths can give rise to undisclosed security vulnerabilities and potentially expose sensitive internal system information. Such vulnerabilities might inadvertently offer attackers opportunities to establish backdoors within the system, as shown in Figure \ref{figure:exa2}(c).  Behavioral benignity authentication uses traceability analysis to locate the minimum-cost repair points for high-risk access paths, achieving proactive prevention. This further enhances the security of intelligent information networks and safeguards the safety of internal system information. Ensuring the legitimacy of access entities, promptly blocking abnormal traffic, and repairing high-risk access paths, behavioral authentication continuously addresses security and safety issues in various aspects of intelligent transportation information services. This plays a critical supporting role in harnessing the efficiency of transportation infrastructure, improving the operational efficiency and management level of transportation systems, and facilitating smooth public travel.

\section{Studies on behavioral authentication}\label{Literature Review}
%\vspace{-0.15in}
%对于行为身份认证而言，设备通常会在后台进行行为数据的收集，然后基于这些手机的数据训练一个机器学习模型，基于训练好的机器学习模型，对后台收集的行为数据进行特征的提取，从而形成用户的行为模板，在进行行为身份认证的过程中，会将新生成的身份凭证与存储的身份信息进行核对（这里加一下形式化的内容），以确定用户身份是否真实，这个过程如图1 所示，也可以结合一个借贷的例子。
%查一下行为身份认证的定义

%行为身份验证技术是通过对用户的行为模式来对用户进行身份验证的一种技术。不同于传统身份验证方法，行为身份验证不需要用户的PIN 码、令牌、短信验证码等数字密码，也不需要用户的指纹、虹膜等生物特征，而是通过用户的持续的、独特的一系列行为特征数据来进行身份验证，如击键行为、触摸手势、行为交互数据等。本文基于前人的工作，将行为特征数据分为6 大类，分别是keystroke、touch gesture、motion、user interaction behavior、intrinsic signaling behavior.
%keystroke指的是用户使用键盘、小键盘甚至鼠标等(keyboard or keypad) 输入的行为，该行为涉及到击键力度、击键速度、击键频率、击键热力图等一系列特征参数。（mouse behavior、keystroke heatmap、keystroke force、typing speed、keystroke frequency） mao 等，shen等 zhu 等，
%touch gesture是指用户使用手势、触摸等方式与设备交互过程中产生的行为，该行为有很多方面的特征，包括finger trajectory、touch pressure、sliding speed、touch area、touch duration、gesture patterns 等等。
%motion是指用户在使用可穿戴式设备或移动设备时，肢体上会做出不同的姿态、动作等行为。这方面的数据主要来自显示人员肢体动作的图片或是由用户设备上的加速度计、陀螺仪等收集记录下的运动数据，因此动作行为具有如下特征：gait、limb、movement、acceleration、orientation 等。
%User interaction behavior是指用户在使用应用程序进行社交、金融交易、浏览信息等交互行为时，产生的一系列交互行为数据，包括但不限于history data、transaction data、sensitive behavior、user profile data、social behaviors。
%Intrinsic signaling behavior是指用户在与设备交互时身体内在的一些信号行为，这些信号行为数据通常需要特定的传感器才能采集，例如EEG、EMG、ECG、breath、expression、voice 等。这些信号行为序列通常也能反应一个人的行为模式。
%multimodal是指在使用多类别的行为数据、或是采取传统方法与行为身份验证相结合的方法对用户进行身份验证。

\subsection{Behavioral identity authentication}
%Behavioral identity authentication refers to verifying users' identity through their behavioral patterns. Unlike traditional identity authentication that requires numerical passwords such as PIN codes, tokens, SMS authentication codes, or biological features like fingerprints or irises, behavioral authentication uses a continuous and unique set of behavioral feature data including keystroke behavior, touch gesture, and behavioral interaction data.

\begin{figure}[t]
  \centering
  \includegraphics[width=1 \textwidth] {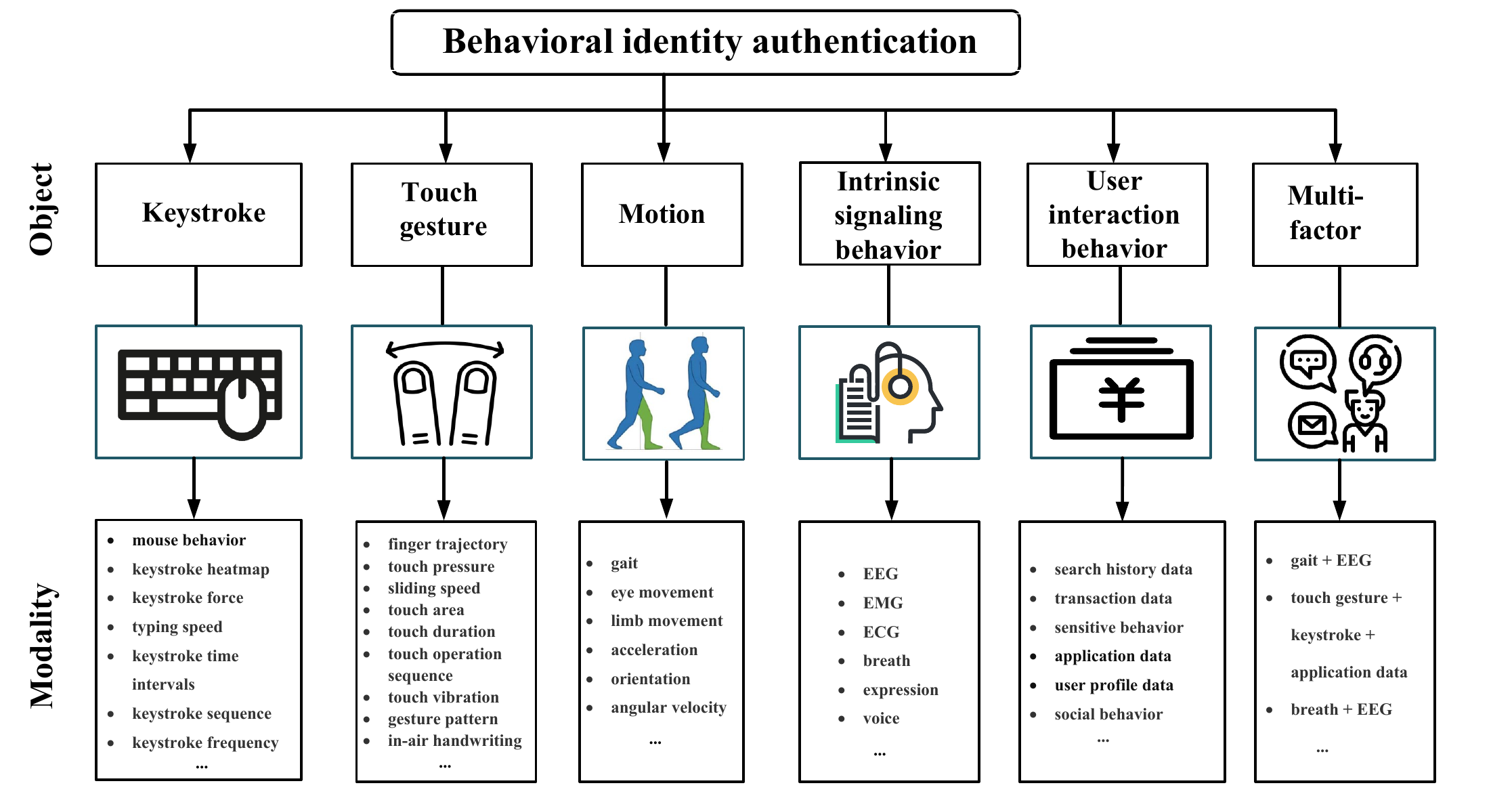}
% \vspace{-0.18in}
  \caption{ Components of behavioral identity authentication. }
  \label{figure:bia}
%  \vspace{-0.1in}
\end{figure}

According to the different characteristics of behavioral data, this part categorizes behavioral identity authentication  into five major types: keystroke, touch gesture, motion, intrinsic signaling behavior, and user interaction behavior. In addition, in certain specific scenarios, the above different behavioral feature types can be combined into multi-factor approach to reflect the user behavior, as shown in Figure \ref{figure:bia}.

 Keystroke-based authentication is one of the earlier behavioral identity authentication methods that verifies a user's identity by analyzing the characteristics of the user's input characters on the keyboard, such as keystroke force, keystroke speed, keystroke frequency, keystroke sequence, and keystroke time intervals. Early research into keystroke-based authentication focused on analyzing keystroke force and frequency. Zhu et al. \cite{zhu2002novel} proposed a novel approach to authenticate users based on keystroke dynamics while entering passwords. The proposed method leverages a keystroke feature vector which consists of the user's keystroke force and frequency to confirm the user's identity. Subsequently, researchers introduced extra keystroke features to continuously improve the accuracy and security of keystroke-based authentication. Primo et al. \cite{primo2017keystroke} investigated the effect of music on keystroke behavioral data using two validation techniques: relative and absolute measurements. The researchers analyzed the data based on three dimensions of keystroke behavior: key hold, key press, and inter-key intervals, to understand how music affects user verification performance. Their findings confirmed that a music environment can improve the accuracy of user authentication based on keystroke dynamics. Ho et al. \cite{ho2018one} introduced a user authentication approach based on keystrokes, which integrates a one-class naive Bayes algorithm (ONENB) and a typing speed inspection in typing skills (SITS) algorithm. The ONENB algorithm computes the probability of keystroke behavioral data, while the SITS algorithm incorporates keystroke patterns to define user keystroke characteristics and builds an authentication model to differentiate legitimate users from impostors. Lee et al. \cite{lee2019parameterized} designed a parameterized model that utilizes a feature selection method based on the median and interquartile range to extract keystroke dynamics features. The model binds the calculated security measure (FAR) obtained from these keystroke dynamics features for user identity verification. Furthermore, with the development of understanding of keystroke dynamics, many input devices have been used for authentication in similar ideas to ensure the security of user's identity, such as mice, and virtual keyboard. Mao et al. \cite{mao2016research} developed an innovative real-time identity authentication technique using mouse behavior learning. This approach entails collecting the mouse behavioral feature vectors of each new user during their initial login, both in dynamic and static scenarios, and it compares them with the feature vectors gathered during subsequent login authentication in simulated abnormal scenarios to attain user identity verification. Shen et al. \cite{shen2012user} proposed a user identification method that involves analyzing mouse behavior. In a controlled environment, data collection is conducted on the mouse operations performed by the user for specific mouse operations. Subsequently, the features are extracted and categorized to accomplish user identity authentication by analyzing the results of the classification. Kang et al. \cite{kang2015keystroke} expanded the applicability of a dynamic user authentication approach that relies on keystroke dynamics. This approach now enables the authentication of users typing various textual strings across multiple input devices while distinguishing between legitimate and potential imposter users. Inguanez et al. \cite{inguanez2016securing} developed a user authentication technique for smart touch screen devices. This method utilizes a multilayer perceptron  (MLP) neural network to classify the graph of the user's keystroke behavior and compares it with the characteristics of the keystroke heatmap in order to achieve user authentication. However, keystroke-based authentication still has certain environmental dependencies and limitations. Variable factors such as different input devices and keyboard layouts may require re-establishing authentication models. The behavioral feature information provided by the keystroke patterns is relatively limited, so more identity authentication methods using other behavioral features have been developed.

 With the popularization of smartphones, touch gesture-based authentication methods have been proposed, which make up for the shortcomings of the insufficient use of behavioral characteristics in the keystroke-based authentication methods. Touch gesture refers to the interaction behavior between users and devices through gestures or touches. Existing authentication schemes usually adopt finger trajectory, touch pressure, sliding speed, touch area, and gesture pattern as behavioral features that need to be extracted and analyzed. Cao et al. \cite{cao2021evidence} devised a continuous user authentication scheme which uses vibration response as an implicit biometric feature. It applies a high-pass filter to remove noise and segments the vibration data for each touch event to achieve accurate data matching and recognition and proposes a novel centroid vector method to infer the touch position accurately. Shen et al. \cite{shen2017performance} proposed a method to use gesture pattern, touch operation sequence, and other behavioral data for identity verification. It uses a Markov classifier to model the motion sensor data sequence and capture the temporal and dynamic features of the sensor events, which improves the security and applicability of identity verification. Mao et al. \cite{mao2022implicit} proposed an implicit continuous authentication model for user authentication on mobile devices through touch behavior. The model uses data from sensors such as accelerometers and gyroscopes to create feature vectors that include both macroscopic and microscopic features. Yang et al. \cite{yang2019behavesense} designed a touch-based behavioral biometric recognition technique using a single-class support vector machine and independent random forest training models. The technique evaluates the accuracy of each type and then applies Bayes' theorem to estimate the confidence level of each type, which provides a safe and continuous authentication method for mobile applications. Xu et al. \cite{xu2020touchpass} combined the physical features of finger touch and the biometric features of touch behavior by a feature fusion authentication framework. It utilizes a particular training sample selection strategy to convert signal features into behavior-agnostic features and subsequently applies knowledge distillation in constructing a touch user authentication scheme. Yang et al. \cite{yang2021enabling} presented an integrated identity verification scheme that combines passwords and touch behavioral factors which include touch pressure and sliding speed. The scheme utilizes a novel algorithm to differentiate fine-grained finger input and supports different forms of passwords in the frequency domain, which improves the security of authentication. The above-mentioned methods, without exception, are only for the authentication of a single user. In actual scenarios, there may be multiple users on the same device. The aforementioned studies struggle to address this type of authentication issue.

 In contrast to keystroke-based and touch gesture-based authentication, motion-based authentication methods emerged later due to the requirement for a multitude of sensors to collect user motion data for analysis. With the development of wearable and mobile devices, more sensors, such as accelerometers and gyroscopes, are built into the devices to meet the needs of users, which also meets the basic conditions of motion-based authentication methods. Existing motion-based authentication schemes typically utilize motion features such as gait, limb movement, acceleration, and orientation to perform identity verification to ensure the security of users' identities. Chen et al. \cite{chen2017modeling} proposed a framework for continuous user authentication based on motion behavioral characteristics, including direction, acceleration, and angular velocity. The framework utilizes smartphones' built-in sensors to collect typical daily motion data of users such as time, frequency, and wavelet domains, which achieves relatively accurate user authentication.
 Lee et al. \cite{lee2016implicit} introduced an advanced smartphone authentication system called iAuth, which leverages the capabilities of multiple sensors embedded within smartphones, Bluetooth connectivity, and wearable devices equipped with sensors. By utilizing machining learning methods to capture the user's unique gait behavior pattern in sensor data from different devices, iAuth enables seamless and ongoing authentication of end-users. Zou et al. \cite{zou2020deep} developed a hybrid deep neural network method for identity verification that extracts robust gait features from time series data. The network combines the convolutional output features with the temporal properties of the data, which enhances the gait features and increases the authentication accuracy. He et al. \cite{he2022gait2vec} developed a gait feature extractor using transfer learning to reduce time costs and and enhance the model's robustness. The extractor has undergone pre-training for user identification tasks. Song et al. \cite{song2016eyeveri} developed an authentication system based on eye movement. The system verifies the user's identity by capturing the features of human eye movement from the front camera of the smartphone. In terms of multi-user authentication, the authentication schemes need to solve the problem of changing motion features to ensure the system's safety. Kong et al. \cite{kong2019gait} presented an optimized method for user identification based on gait features. To accurately capture the behavioral characteristics of gait data, their method employs a spatial transformation algorithm to optimize coordinate drift and utilizes a support vector machine algorithm to address the issue of gait feature changes when switching between users. Compared to keystroke-based authentication methods, motion-based authentication methods involve multiple sensor data that may be leaked during transmission, thereby compromising the safety of the system. Wang et al. \cite{wang2021framework} proposed a novel behavioral authentication framework for user motion characteristics to address issues including behavioral dynamics, data privacy, and side-channel leakage. The framework accelerates feature transfer speed on mobile devices and mitigates potential side-channel leakage, and improves security during transmission. In summary, motion-based authentication schemes introduce more sensors to capture behavioral features, which makes full use of user behavioral features but also brings greater computational overhead and sensor data leakage problems that necessitate optimization and resolution.

 Compared with touch gesture and motion, intrinsic signals, such as EEG, EMG, ECG, breath, facial expression, voice, and others, are highly unique in behavioral recognition, making it more difficult for attackers to accurately replicate. Moreover, existing research also focuses on these sequences of behavioral signals which typically reflect an individual's behavior patterns. Chauhan et al. \cite{chauhan2020contauth} proposed the ContAuth system, which targets inherent behavioral signals of users, such as breath and EEG, obtained from sensors using a class-incremental learning method. It combines deep learning models with online learning models to enhance the robustness of behavior-based authentication. Perera et al. \cite{perera2018face} designed a sparse representation-based multi-user mobile active authentication scheme according to the dynamic facial expressions of users that automatically adjusts the parameters using the extremum distribution mechanism. It also includes an extension algorithm for applying the scheme in a single-user scenario. Lu et al. \cite{lu2018lippass} studied the acoustic Doppler effect of user speech and created a lip-reading-based user authentication system called LipPass that operates in noisy environments. Ji et al. \cite{ji2020nonlinearity} introduced a position-sensitive identity verification mechanism called NAuth with nonlinear enhancement. This verification scheme ensures consistent device identity verification by extracting acoustic nonlinear patterns (ANP). Implementing intrinsic signal authentication often requires specialized hardware, which may not be available on all devices. Also, acquiring these signals often requires stable physical and emotional states, and changes in signals caused by emotions or illness may affect the performance of the system.

 In addition to the above-mentioned methods relying on devices or sensors for identity authentication, some methods can achieve identity authentication only through user interaction data. With the advancement of big data, machine learning, artificial intelligence, and other technologies, these methods have received more and more attention and research. User interaction behavior, such as social networking, financial transactions, and information browsing has generated a series of data that includes but is not limited to search history data, transaction data, user profile data, sensitive behaviors, and social behaviors. Ruan et al. \cite{ruan2015profiling} introduced a user authentication scheme that relies on social behavioral characteristics. They extracted and classified social behavioral features such as user browsing and clicking on OSN websites, determined metrics for each feature, and constructed user behavior profiles. Finally, they validated the accuracy of distinguishing genuine users from impostors using user behavior profiles. Shi et al. \cite{shi2011implicit} presented an implicit identity verification scheme based on user behavior patterns, leveraging history data on smartphones and movement data collected by sensors to extract behavioral features for user authentication. Skravcic et al. \cite{skravcic2017authentication} presented an implicit authentication scheme  centered around user behavior patterns. This approach employs classification models that are built using vast amounts of user transaction data, call records, and email correspondences from various systems, such as banking and social networks, to identify legitimate and malicious users. By leveraging these behavior patterns, the scheme effectively distinguishes between these two user categories. Yang et al. \cite{yang2017energy} designed a wind vane module to achieve lightweight implicit authentication. This module determines the amount of data needed to be collected at different times based on the user's identity legitimacy and interactive behavior habit and adjusts the sampling rate accordingly, which provides an energy-efficient solution for real-time implicit authentication on mobile devices. Shi et al. \cite{shi2022identity} developed an end-point identity authentication technology based on the analysis of user-associated behaviors. The approach employs an interactive behavior common-subsequence similarity algorithm, which extends the traditional behavior common-subsequence (BCS) sequence pattern, and considers the maximum overlap of user behavior sequences and the short sequence overlaps at different time intervals to better identify any anomalies during each user's login session. Wu et al. \cite{wu2021network} proposed a Hidden Markov Model (HMM) to detect malicious user interactive behavior in network systems by extracting relevant features and defining the observation symbols and hidden states based on the user's access behavior patterns. It maps the user's behavior to an HMM chain and identifies any abnormal or harmful actions to ensure the security of network systems. Such methods need to collect, analyze, and store user interaction behavioral data, especially for behaviors containing sensitive data, which may involve privacy issues. Furthermore, such methods are prone to adversarial attacks. When the attacker adds fake historical data to the training samples, the model will not be able to perform correct authentication.

\begin{table}[t]\scriptsize
\centering
        \caption{Summary of  behavioral identity authentication}
    %    \begin{tabular}{>{\centering\arraybackslash}m{2.7cm} p{4.1cm} p{4.0cm} p{3.5cm} }
     \resizebox{1\textwidth}{!}{

\begin{tabular}{m{2.8cm}<{\centering}m{4.1cm} m{4.1cm} m{3.5cm} }
        \toprule[\heavyrulewidth] % Top line with heavier thickness
    \multicolumn{1}{c}{\textbf{Object} }&\multicolumn{1}{c}{\textbf{Description}}&
    \multicolumn{1}{c}{\textbf{Characteristics}}& \multicolumn{1}{c}{\textbf{Issues}}\\
       \midrule
   Keystroke \cite{lee2019parameterized,zhu2002novel,primo2017keystroke,ho2018one,mao2016research,shen2012user,kang2015keystroke,inguanez2016securing} & Keystroke refers to the action of a user inputting information through a keyboard, a keypad, or even a mouse. &  (1) Without reliance on additional devices. (2) Extra layer of security for password. (3) Non-invasive continuous authentication. & Data quality easily affected by environmental and user conditions. \\
        \midrule % Line to separate the Goal section
      Touch gesture
      \cite{xu2020touchpass,yang2019behavesense,cao2021evidence,shen2017performance,mao2022implicit,yang2021enabling} & Touch gesture refers to the interaction between users and devices through gesture or touch. &  (1) Covering multiple dimensional features to improve the security of verification.  (2) Ensuring frictionless authentication. &  (1) Difficulty in distinguishing and identifying multiple behavior patterns. (2) Cross-platform versatility limited by inconsistent behavioral data formats. \\
        \midrule % Line to separate the rest of the rows
Motion \cite{he2022gait2vec,wang2021framework,zou2020deep,chen2017modeling,lee2016implicit,song2016eyeveri,kong2019gait} & Motion refers to the various postures and movements made by users while using wearable or mobile devices.  & (1) Utilization of broader behavioral features. (2) Non-invasive continuous authentication. (3) Less prone to variation caused by external factors. & (1) Multi-device dependency. (2) Privacy risks posed by sensor attacks. \\
        \midrule % Line to separate the rest of the rows
Intrinsic signaling behavior \cite{perera2018face,chauhan2020contauth,lu2018lippass,ji2020nonlinearity} & Intrinsic signaling behavior refers to the signals emitted by human organs during the interaction between users and devices. &  High biometric uniqueness.  & (1) High device dependency. (2) Data quality largely affected by emotions, diseases, etc.\\
          \midrule % Line to separate the rest of the rows
      User interaction behavior \cite{yang2017energy,shi2022identity,wu2021network,ruan2015profiling,shi2011implicit,skravcic2017authentication} & User interaction behavior refers to the behavior of users interacting with applications. & (1) No requirement for sensor data collection and conversion. (2) A wider range of behaviors beyond keystroke or motion. & (1) Suffering from data privacy problem and breach risk. (2) The increasing probability of misjudgments of the model due to adversarial attacks.\\
          \midrule % Line to separate the rest of the rows
Multi-factor \cite{zhang2020deepkey,dasgupta2016toward,wazzeh2022privacy,liu2018multi}& Multi-factor refers to the use of multiple categories of behavioral data or a combination of conventional methods with behavioral authentication to verify a user's identity. & (1) Flexible combination of authentication methods. (2) The addition of cross-validation for an extra layer of security. (3) Reducing reliance on a single piece of sensitive data. & (1) The challenge of seamlessly integrating various behavioral features and authentication technologies. (2) Imbalanced data from various behavioral features. \\
        \bottomrule[\heavyrulewidth] % Bottom line with heavier thickness
    \end{tabular}}
    \label{sumbia}
    \vspace{0.1in}
\end{table}
 With the further development of authentication technology, some studies realized that multi-factor authentication systems, which rely on multiple factors to provide robust and accurate results, have stronger security compared with systems that only consider a single behavioral feature. Dasgupta et al. \cite{dasgupta2016toward} proposed a multi-factor authentication system that considers various combinations of different data features through a subset of available authentication modalities, which ensures efficiency in dynamic environments. They conducted tests on one-time and continuous authentication for smartphone users and confirmed that there is complementarity between different signals, which can enhance the performance of the authentication system. Zhang et al. \cite{zhang2020deepkey} designed a multi-modal biometric authentication system that combines EEG and gait data and leverages their unreplicated characteristic. This system uses a doubly authenticated method to improve the anti-counterfeiting and security of the authentication process. Wazzeh et al. \cite{wazzeh2022privacy} devised an authentication scheme for mobile devices utilizing  federated learning (FL). This scheme allows each user to keep their private data locally for safety and trains models to capture their multi-modal behavioral data with a server for global aggregation. Liu et al. \cite{liu2018multi} presented a user authentication method for smartphones by analyzing user interaction behavior. The approach establishes a behavioral characteristic classification model using data such as the user's touchscreen interaction method, motion, and phone power consumption to enable continuous user identity verification. However, when designing multi-factor authentication methods, suitable algorithms are needed to solve problems such as multi-factor data fusion and data imbalance, and it is technically challenging to seamlessly integrate various factors.

 Based on the aforementioned analysis, characteristics and issues of different objects of behavioral identity authentication are summarized in Table \ref{sumbia}.

\begin{figure}[t]
  \centering
  \includegraphics[width=1 \textwidth] {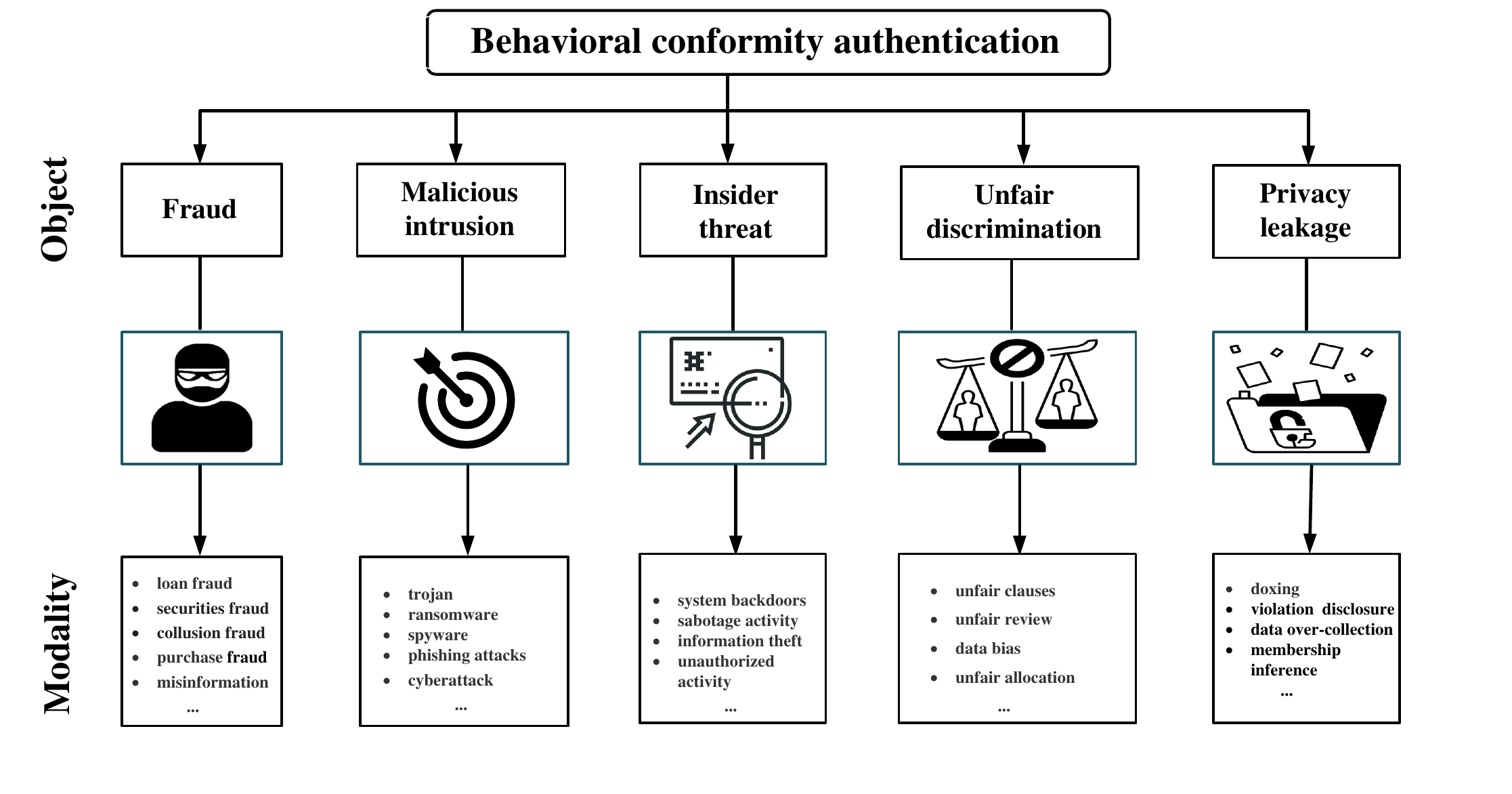}
% \vspace{-0.18in}
  \caption{ Components of  behavioral conformity authentication. }
  \label{figure:bca}
 \vspace{0.05in}
\end{figure}
\subsection{Behavioral conformity authentication}

Behavioral conformity authentication aims to identify potential security and safety risks within a system that fall outside the scope of behavioral identity authentication. The risky objects detected by behavioral conformity authentication mainly include five  typical categories, i.e., fraud, malicious intrusion, insider threat, unfair discrimination, and privacy leakage, as shown in Figure \ref{figure:bca}.

     Fraud risks commonly occur in industries such as telecommunications, healthcare, and finance, with a particularly significant impact observed in the financial field. In the financial domain, user identity authentication is typically subjected to heightened scrutiny. This is attributed to the involvement of substantial funds and sensitive information in financial transactions, necessitating the assurance of transactional security and accuracy. Despite the stringent measures, instances of loan fraud still occur, even when identity authentication processes are in place. For fraudulent behavior in the financial sector, some methods can detect non-compliant activities and transactions by monitoring customer transactions and behavior patterns. Jiang et al. \cite{jiang2021financial} developed a more comprehensive network for embedding location information, called the Fuller Location Information Embedding Network. The network employs self-supervised learning to characterize the address features of users, with a focus on analyzing the relationship between address information, behavioral information, and customer fraudulent behavior in loan applications, effectively improving the performance of loan fraud detection models. Wu et al. \cite{wu2022fraud} proposed a two-stage detection model for identifying fraudulent agents on online large-scale loan platforms. The model extracts $26$ features from activity records such as communication logs and application activity histories of agents and borrowers, as well as loan histories. Based on these $26$ features, the model characterizes the behavior patterns of the classification objects in detail and achieves high accuracy in identifying fraudulent agents. Awotunde et al. \cite{awotunde2022artificial} proposed an artificial neural network-based detection method for the fraud in bank loan management. They extracted a series of borrower-related and loan-related information data as features for identification and classification, in order to detect fraudulent behavior in loan transactions. Chang et al. \cite{chang2022design} presented a general model for detecting financial fraud using natural language processing technology to accurately detect and classify fraud. As an instance of the model, they implemented an anti-fraud chatbot on a widely used social network service. Nevertheless, certain advanced fraudulent activities transcend the actions of solitary individuals, involving collaboration among multiple users or intermediary agents. In view of these scenarios, a number of research studies have been dedicated to investigating the correlations within the behaviors of multiple users. Xu et al. \cite{xu2021towards} proposed a novel graph neural network with a role-constrained conditional random field (GRC) for loan fraud detection. The model utilizes graph neural network to detect individual user loan fraud and collusion fraud based on borrower role information and network social relationships.  Experimental results demonstrate that the model performs well in detecting loan fraud on Alipay. Wang et al. \cite{wang2022wrongdoing} proposed a graph-based approach for behavior modeling called behavior identification graph (BIG). This method delves into the property-level associations in behavioral data and integrates the inter and intra behavioral correlations into a unified space. Furthermore, they introduced a property graph to describe fine-grained correlations between properties, where the structure of the graph corresponds to the topology information of behavior events. Based on the  property graph, they designed an event-property composite model and used network representation learning algorithms to extract fine-grained associations at the behavioral property-level. The behavior patterns are represented in a multidimensional spatial distribution of behavioral properties. The effectiveness of this method has been verified in various network threat detection scenarios, particularly in the fraud detection scenario. In addition, some research has focused on purchase fraud and misinformation. He et al. \cite{he2021datingsec} proposed a novel malicious user detection system, called Datingswc, that aims to address fraudulent activities such as misinformation or illegal information on dating applications. The system utilizes user profiles, comments, and other relevant information to establish behavior patterns using machine learning models, such as MLP and LSTM. These behavior patterns are then combined and inputted into an attention module to automatically detect suspicious activities following these behavior patterns. Wang et al. \cite{wang2023purchase} designed a novel graph-based fraud detection framework to detect fraudulent orders placed by employment fraud teams. The framework comprises two parts: the DPP module, which extracts feature sequences of user click locations from the website, and the GSR module, which performs neighborhood sampling and information aggregation. The effectiveness of this framework has been validated through purchase fraud detection on the JD platform.  Wang et al. \cite{wang2021ifdds} developed an interactive fraud detection dialogue system, which actively engages in conversations with clients by means of intelligent voice interactions. The system employs imitation learning to master the dialogue strategy and accurately assesses the risk of actual payment, thereby reducing the likelihood of misjudging payment behavior without actual risk. However, behavior-based fraud detection methods still exhibit certain shortcomings and face a set of challenges. Constructing a fraud model necessitates the inclusion of user privacy data, which requires the consent and authorization of both the regulatory platform and the user. Once privacy data is acquired, the model confronts the difficulty of addressing the inherent imbalance between instances of fraudulent and normal behavior during the training phase. Furthermore, the trained model must contend with the ever-evolving deceptive techniques, thereby necessitating regular updates to accommodate the identification of emerging fraud patterns.

    Different from fraud with deceptive features, malicious intrusion has evident attack-oriented features. This type of behavior poses substantial risks, not only to personal computers but also to smartphones, Internet of Things (IoT) devices, and even to network security as a whole. The category of malicious intrusion spans a broad spectrum, including trojans, ransomware, spyware, phishing attacks, and cyberattacks. Existing intrusion detection frameworks often utilize machine learning algorithms to extract features from network behavior, aiding in the differentiation between normal and malicious activities. Chen et al. \cite{chen2022efficient} presented an efficient Network-Based Anomaly Detection (NBAD) algorithm using a combination of Deep Belief Network (DBN) and Long Short-Term Memory (LSTM) network for cyberattack detection. Firstly, on the premise of maintaining the accuracy, the DBN method is utilized to automatically extract the features of the original data nonlinearly, so as to express the features of the original data with a lower dimension. Furthermore, the classification results, as the basis for identifying anomalous network behavior, are obtained through a lightweight LSTM network. Chen et al. \cite{chen2019automated} presented an automated ransomware pattern extraction and early detection tool.
    The tool analyzes discovered malware samples and generates a log, from which it extracts sequences of events triggered by ransomware. It also ranks the features of ransomware and detects malicious activity from the learned behavior, which effectively enhances the security of infrastructures.
     Hamid et al. \cite{a2011hybrid} designed a hybrid feature selection method for detecting phishing attack behavior, which combines content-based and behavior-based analysis. The method analyzes the content and ID tags of phishing emails to identify characteristics of attacker behavior and subsequently detects phishing attacks. Qin et al. \cite{qin2022interaction} devised a novel unsupervised network behavior anomaly detection framework, which combines real-time high-order host state in a dynamic interactive environment with dialogue patterns between hosts. It automatically generates high-order features from a series of basic features extracted from the graph neural network (GNN) and identifies various cyberattack behaviors more effectively. Jiang et al. \cite{jiang2020intelligent} proposed a behavior-based method for intelligent recognition and security supervision of unmanned aerial vehicles (UAVs). The method uses location tracking and flight data from the onboard black box to collect real-time behavioral data of UAVs. Then, it identifies suspected intrusion and attack behaviors of UAVs through behavior modeling and issues warnings in potential illegal situations. Garg et al. \cite{garg2019hybrid} addressed the issue of high false alarm rates in existing real-time anomaly detection by proposing a hybrid detection model that utilizes grey wolf optimization (GWO) and convolutional neural networks (CNN). The model improves the feature selection and anomaly classification capabilities. In the first stage, improved grey wolf optimization (ImGWO) is used for feature selection to minimize the feature set. In the second stage, optimized convolutional neural network (ImCNN) is used for more effective anomaly classification. Pajouh et al. \cite{pajouh2016two} devised a novel intrusion detection model that uses two layers of dimensionality reduction and two layers of classification modules to identify malicious activities such as User to Root and Remote to Local attacks. The model uses Naive Bayes and the Deterministic-KNN version to detect suspicious behavior. Wang et al. \cite{wang2020iot} proposed a security detection system (IoT-Praetor) for malicious attacks on IoT devices. The system uses a novel DUD model to construct norms for the interaction and communication behavior of IoT devices. A behavior rule engine is employed to detect device behavior in real-time, enabling the identification of behaviors that damage devices through malicious network communication. Some studies also consider the security of data privacy and the safety of the detection framework while striving for accurate detection of malicious behavior. Pei et al. \cite{pei2022personalized} devised a personalized federated anomaly detection framework in order to take into account privacy protection in the process of detecting anomalous network traffic. It archives a personalized detection model by fine-tuning the model structure of different systems, which improves the data utility on the premise of protecting the privacy and takes into account the safety of the method while improving efficiency. Mothukuri et al. \cite{mothukuri2021federated} presented a federated learning-based method for detecting malicious attacks in IoT. The method utilizes decentralized device data to proactively identify intrusion behaviors in IoT networks, achieves privacy preservation of terminal devices, and performs better in attack detection than non-federated learning methods. Kurt et al. \cite{kurt2022online} devised a data-driven method to detect UDP flooding and spam attacks in IoT networks. To ensure the privacy of node data, the scores are encrypted and perturbed before being sent to the network operator for aggregation of statistical information, followed by anomaly detection through generalized accumulation and algorithms. Intrusion detection methods still have some deficiencies and challenges. Similar to fraud detection, the model of intrusion detection also has the problem of imbalanced training data. When the intrusion detection model needs to be deployed on multiple devices, the problem of heterogeneous data fusion is also considered.

    The above-mentioned fraud and malicious intrusion are both behaviors that occur outside the system, while insider threats are relatively intuitive risk behaviors inside the system, which may also lead to system collapse or huge economic losses. System backdoors and equipment failures are relatively common insider threats. Ji et al. \cite{ji2022proactive} designed an active anomaly detection network for mobile robots to address system malfunctions caused by outdoor environmental factors. The network effectively integrates multiple sensor signals to ensure robust anomaly detection even in the presence of sensor obstruction in the field environment. Cui et al. \cite{cui2021security} introduced a blockchain-supported decentralized and asynchronous FL framework for anomaly detection in IoT systems. This framework ensures data integrity, avoids single-point failures, and improves the security of the system. Luo et al. \cite{luo2018distributed} first introduced autoencoder neural networks to solve anomaly detection problems in wireless sensor networks. By constructing a three-layer autoencoder neural network, they overcame the huge demands for network resources that deep learning requires. This method is mainly used to detect IoT device failures and changes in the environment. Li et al. \cite{li2022situation} proposed an active learning and contrastive-based detection model, which monitors the system performance indicators such as CPU utilization, latency to form a multivariate time series. It models the abnormal and normal sequences based on the VAE model to recognize the abnormal sequences and discover the potential security backdoors in the system. In addition, information theft, unauthorized activity, and even sabotage activity by system insiders are also insider threats. Modell et al. \cite{modell2021graph} devised a incremental approach to analyze anomalous user behavior in event logs. It employs graph embedding to acquire a vector representation of the users, which is updated over time and utilized to model the configuration profiles of user-accessed resources and builds the formation of a dynamic, interactive network comprising users and resources. The method is applicable for identifying suspicious or unauthorized user behavior in enterprise networks. Hou et al. \cite{hou2022lightweight} proposed a lightweight framework for detecting abnormal driving behavior. The framework utilizes IoT devices as carriers and captures video and image data using cameras. Based on the analysis of this data, the framework identifies dangerous behaviors such as fatigue driving and erratic driving. Mazzawi et al. \cite{mazzawi2017anomaly} proposed a machine learning algorithm for detecting malicious user activity in databases, which is used to detect suspicious behaviors, such as information theft and unauthorized activity. The algorithm consists of two primary components: one is responsible for generating models using user behavior, and the other focuses on clustering similar behaviors to detect abnormal patterns that might be shared among a group of users. The current detection methods rely on high-quality behavioral data to ensure models' performance. When encountering sporadic zero-day backdoors or highly covert  unauthorized activities, these existing detection methods have difficulty adapting rapidly to newly emerging threats due to a lack of behavioral data.
   % Apart from relying on historical data for modeling, current detection methods also depend on predefined rules or patterns, which necessitate substantial expert knowledge. However, when confronted with zero-day backdoors or highly covert unauthorized behaviors, these existing detection methods struggle to promptly adapt to emerging threats.

    Unlike insider threats, the risk of unfair discrimination might not be immediately apparent, but rather, it accumulates gradually within the system over time. For instance, within certain machine learning models, data bias could propagate unfairness to model decisions. In cloud environments, resources might be unfairly distributed due to scheduling or allocation mechanisms. If the resources allocated to maintain the security of the environment fall short, the environment will gradually fall into a perilous situation. Additionally, on some network platforms, there may be unfair user clauses. Existing research mainly focuses on detecting or optimizing problems such as data bias, imbalanced distribution, and unfair clauses. Li et al. \cite{li2019fair} designed a q-Fair Federated Learning (q-FFL) optimization method that enables fairer performance allocation among devices in large-scale federated networks. This method can also be applied to other related problems such as meta-learning, which helps in fair initialization across multiple tasks. Mohri et al. \cite{mohri2019agnostic} developed a novel unbiased Federated Learning framework to address the issue of model bias among different clients in Federated Learning scenarios. This framework is also applicable to learning scenarios such as cloud computing, domain adaptation, and data drift. Wei et al. \cite{wei2010game} designed a method for optimizing resource balance distribution in cloud services. First, they employed a binary integer programming approach to address the resource allocation optimization problem among independent applicants. Second, they used evolutionary programming to modify the reuse strategy of initial optimal solutions of different applicants. This method provides a solution for resolving the complex issue of resource balance distribution in cloud computing. Lin et al. \cite{lin2018fair} proposed a combined approach of single-layer dominant and max-min fair (SDMMF) allocation and multilayer dominant and max-min fair (MDMMF) allocation to address the issue of fair resource allocation in Intrusion Detection Systems (IDS) in edge computing. The IDS architecture is divided into six layers, and SDMMF allocation is executed recursively starting from the first layer until resources are assigned to the bottom layer, resulting in the equitable allocation of resources that achieves both single-layer and multi-layer fairness in terms of multiple resources. Lippi et al. \cite{lippi2019claudette} proposed a machine learning and natural language method for detecting unfair clauses in applications or websites. They defined unfair terms and expanded the corpus of clauses in order to better train the model. In addition, the model can not only perform classification tasks but can also identify more information in the clauses and detect and classify the implicit unfair semantics of clauses in the terms. Dolly et al. \cite{dolly2022unfair} designed a scheme for detecting unfair reviews. This scheme uses sentiment analysis algorithms and supervised techniques to determine the overall semantics of customer comments based on the positive or negative emotions reflected in them. All of the above detection or optimization methods do reduce the unfairness of the target, but the selection of fairness indicators, such as $\alpha$-fairness\cite{li2019fair}, has certain subjectivity. The fairness indicators vary depending on different models and scenarios. Therefore, universally applicable fairness indicators should be further studied to promote the formulation of universal fairness clauses.
\begin{table}[t]\scriptsize
    \centering
       \renewcommand{\arraystretch}{1.7}
        \caption{Summary of  behavioral conformity authentication}
         \resizebox{1\textwidth}{!}{
        \begin{tabular}{m{2.7cm}<{\centering}m{4.2cm} m{4.0cm} m{3.4cm} }
        \toprule[\heavyrulewidth] % Top line with heavier thickness
    \multicolumn{1}{c}{ \textbf{Object} }&\multicolumn{1}{c}{\textbf{Description}}& \multicolumn{1}{c}{\textbf{Characteristics}}& \multicolumn{1}{c}{\textbf{Issues}}\\
        %\begin{tabular}{p{3.0cm} p{4.0cm} p{4.2cm} p{3.5cm} }
%\begin{tabular}{>{\centering\arraybackslash}p{3.2cm}>{\centering\arraybackslash}p{4.5cm}>{\centering\arraybackslash}p{2.6cm}>{\centering\arraybackslash}p{2.6cm}}
        \midrule
      Fraud \cite{wang2021ifdds,wang2022wrongdoing,jiang2021financial,wu2022fraud,awotunde2022artificial,chang2022design,xu2021towards,he2021datingsec,wang2023purchase} & Fraud refers to the behavior of fraudsters for obtaining illegal benefits through fraudulent means, such as  concealing facts and stealing information.   & (1) Ensuring process security by detecting suspicious activity during transactions. (2) Uncovering hidden fraud networks through behavioral correlation analysis. & (1) Possible leakage and illegal use of user privacy data. (2) Imbalanced training data. (3) Hard to deal with new fraud patterns timely. \\
        \midrule % Line to separate the Goal section
Malicious intrusion \cite{kurt2022online,garg2019hybrid,chen2022efficient,chen2019automated,a2011hybrid,qin2022interaction,jiang2020intelligent,pajouh2016two,wang2020iot,pei2022personalized,mothukuri2021federated}  & Malicious intrusion refers to attacks from outside the system, posing threats to personal computers, mobile phones and even the entire network. & (1) Ensuring system security by detecting malicious activity. (2) Timely alerts and responses. & (1) Difficulty in detecting advanced persistent threats.  (2) Imbalanced training data.\\
        \midrule % Line to separate the rest of the rows
Insider threat
\cite{hou2022lightweight,ji2022proactive,cui2021security,luo2018distributed,li2022situation,modell2021graph,mazzawi2017anomaly} & Insider threat refers to the risks existing in the system, affecting the safety of the system from the inside. & (1) Enhancing network safety and system availability. (2) Strengthening the compliance of systems, such as enterprise networks. & (1) Dependencies on high-quality behavioral data. (2) Difficult to address zero-day  backdoors. \\
        \midrule % Line to separate the rest of the rows
 {\textbf{}} Unfair discrimination
 \cite{li2019fair,mohri2019agnostic,wei2010game,lin2018fair,lippi2019claudette,dolly2022unfair} & Unfair discrimination refers to unfairness or imbalance within a system, which gradually become apparent as the system runs. & Detecting patterns of unfair discrimination at an early stage allows system personnel to intervene and correct such practices timely, which avoids the cumulative risk of unfair discrimination. & The scarcity of data complicates the comprehensive selection of fairness indicators and hinders the establishment of universally applicable fairness metrics.\\
          \midrule % Line to separate the rest of the rows
 Privacy leakage
 \cite{li2015iccta,liu2017location,mehdy2021multi,shokri2017membership,karimi2022automated,chen2023opportunity,rahat2021automated,li2015privacy} & Privacy leakage usually refers to the exposure of users' private data due to security issues, but it also includes disclosure by a second party, a third party, or even other users. & (1) The discovery of external threats and enhancement of system security through privacy leak detection. (2) The improvement of privacy safety through privacy protection measures. & Under data constraints, such as scenarios where there are no clearly defined
levels of privacy data, establishing an efficient detection model is a subsequent challenge. \\
        \bottomrule[\heavyrulewidth] % Bottom line with heavier thickness
    \end{tabular}}
        \vspace{0.1in}
    \label{sumbca}
\end{table}

    With the emergence of privacy protection standards such as CCPA \cite{stallings2020handling} and GPDR \cite{godinho2021consumer}, people pay more attention to the security of private data. Privacy leakage has become a prevalent security concern. Specifically, privacy leakage may arise not only due to malicious intrusion or insider threat but also user disclosure. All these activities will lead to the result of system or user privacy leakage. Current research predominantly centers on two key aspects: the detection of privacy leakage and the enhancement of privacy protection measures. In terms of detection, several studies have explored issues related to privacy leakage. Li et al. \cite{li2015iccta} proposed a static taint analyzer for detecting sensitive data leaks during data propagation between application components. This method focuses on contextual information regarding data propagation between multiple components, which allows for higher performance. Liu et al. \cite{liu2017location} conducted a study on the problem of privacy leakage caused by background access to location in location-based service applications. Their research shows that accessing locations in the background of an application can generate user movement trajectory data, thereby identifying personal information and resulting in privacy leakage. Mehdy et al. \cite{mehdy2021multi} proposed a hybrid neural network model with multiple inputs and outputs for detecting privacy leaks. The model incorporates pre-trained language models, semantic analysis, linguistics, and other knowledge to accurately identify personal information related to health, finance, and social relationships that Twitter users may disclose when posting tweets. This enables the detection of privacy disclosures by users. Shokri et al. \cite{shokri2017membership} conducted a study on the privacy implications of machine learning models, specifically focusing on membership inference attacks that can lead to data leakage. They conducted experiments using a dataset that contained hospital-related information and analyzed various factors that influence the privacy leakage. Additionally, they evaluated different strategies to mitigate these privacy risks. Karimi et al. \cite{karimi2022automated} proposed an automated method for detecting privacy violations on Twitter. The approach utilizes contextualized string embedding to detect sensitive information in tweets, specifically targeting second-party and third-party doxing and malicious information disclosure, while excluding instances of self-disclosure by users and privacy disclosures not targeting any specific identity. In addition, some researches are dedicated to the improvement of privacy protection. Wang et al. \cite{chen2023opportunity} developed a new solution for the challenges faced in the collection of multidimensional data under Local Differential Privacy (LDP), including high communication costs and noise issues. The solution includes a Multivariate k-ary Randomized Response (kRR) mechanism called multi-KRR to reduce communication cost, a Markov-based dynamic privacy budget allocation mechanism called Markov-kRR to mitigate the impact of noise, and an improvement on Markov-kRR flipping times threshold to optimize data utility. Rahat et al. \cite{rahat2021automated} developed a convolutional neural network-based privacy policy classification model to assess compliance with the privacy policies of various websites. They used General Data Protection Regulation (GDPR) as a standard and identified $18$ labels from it to annotate and classify the privacy policy dataset. Their experiments demonstrated that very few source websites in the dataset strictly followed the GDPR standards. Li et al. \cite{li2015privacy} proposed a mobile cloud framework to prevent applications from over-collecting user data. This framework stores all user data in the cloud and restricts application access to user data in the cloud, proactively eliminating instances of data over-collection. However, different application services have different protection levels for private information, and the existing privacy leakage detection models are only established for specific scenarios and lack of sufficient scenario transferability. Confronted with scenarios where there are no clearly defined levels of privacy data, the detection model may struggle to yield satisfactory outcomes. Establishing an efficient detection model under such data constraints is a subsequent challenge.

  From the above analysis, characteristics and issues of different objects of behavioral conformity authentication are summarized in Table \ref{sumbca}.

\subsection{Behavioral benignity authentication}
\begin{figure}[t]
  \centering
  \includegraphics[width=1 \textwidth] {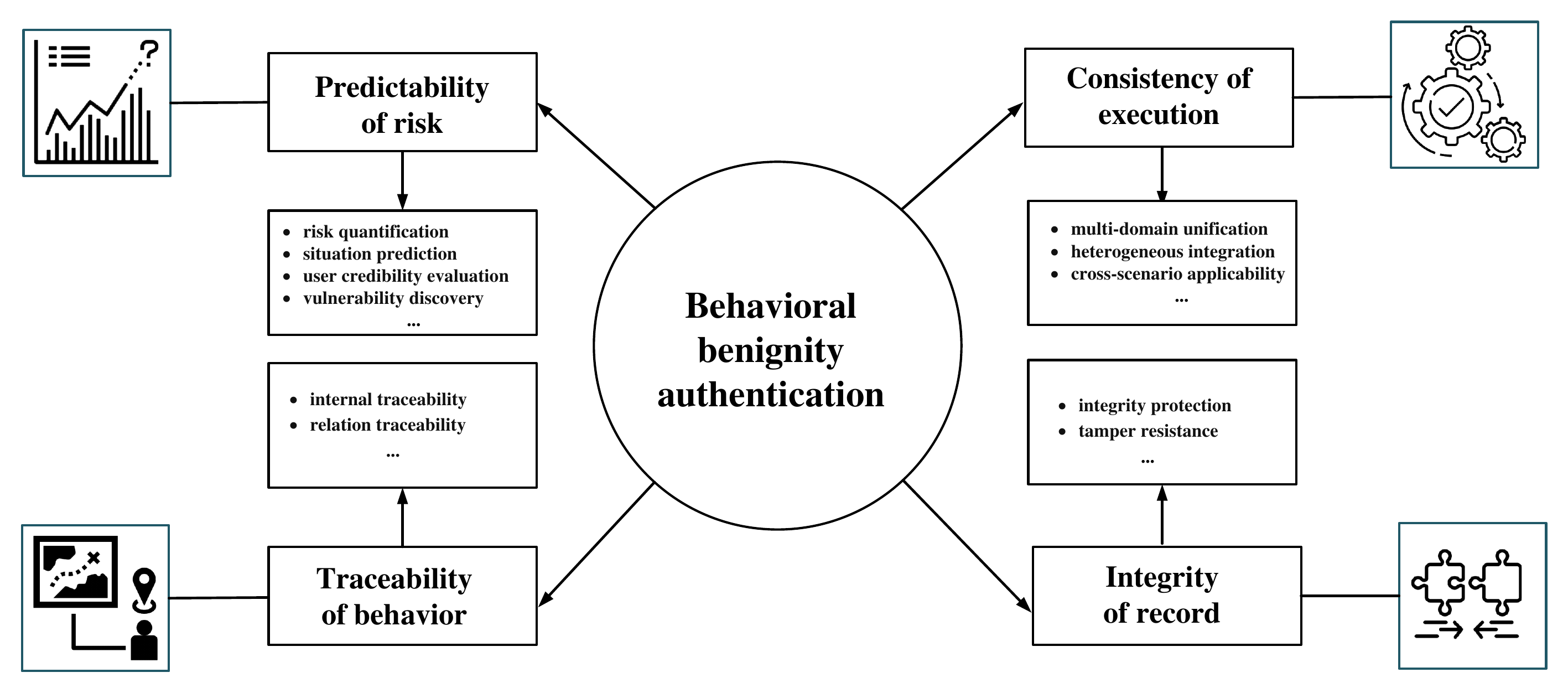}
% \vspace{-0.18in}
  \caption{ Components of behavioral benignity authentication. }
  \label{figure:bba}
%  \vspace{-0.1in}
\end{figure}

The study of behavioral benignity authentication can be mainly divided into the following four aspects: predictability of risk, consistency of execution, traceability of behavior, and integrity of record, as shown in Figure \ref{figure:bba}.

    Predictability of risks is one of the important attributes of a secure network or system. By analyzing and assessing potential risks, the system can identify possible threats and vulnerabilities in advance and take preventive measures.
%    Predictability of risks is one of the important attributes of a secure network or system. Predicting risks serves to prevent potential attacks and exploitable vulnerabilities, aiding in the efficient allocation of resources by directing more towards potentially high-risk areas, thereby ensuring the security of the system.
     Even in cases where risk threats cannot be entirely mitigated, early warnings can significantly reduce the impact of risk events. Many studies, aimed at risk prediction, have introduced methodologies encompassing risk quantification and credibility assessment. Hu et al. \cite{hu2018security} proposed a new threat identification method and risk quantification model for  predicting threats in multimedia communication networks. The threat recognition method uses a dynamic Bayesian attack graph based threat prediction algorithm, which aims to predict threat scenarios using complete information. The risk quantification model quantifies the risk status of the entire network and individual hosts by analyzing the security risks at the host and network levels. Wang et al. \cite{wang2023research} proposed a network behavior risk measurement method by analyzing network traffic. They considered traffic data as network behavior and characterized the network traffic and network topology information. Additionally, they introduced the theory of differential manifolds to measure the behavior risk of the network system. Li et al. \cite{LiQiao2014}  developed a trust-based model for detecting suspicious behavior in network groups. Firstly, the model constructs a trust matrix between network nodes utilizing network topology information. Secondly, it calculates the similarity matrix of nodes based on their trust levels. Finally, the model clusters the nodes utilizing the similarity matrix to identify potential malicious groups. This approach has a high distinction rate for hidden risks in the network. There are also some studies that indirectly conduct risk prediction through network situation prediction methods. Yang et al. \cite{yang2022network} proposed a network security situational assessment method based on attack intent distinction. They analyzed the correlation between the attack phase, network configuration information, and attack intent. The method distinguishes the attacking intent and predicts the next attack, which does not rely on historical sequences and is more effective in predicting network security situational assessments. Ghazel et al. \cite{ghazel2009using}  proposed a global model involving LC (Logistics Center) region's railway and road transportation, where each model describes the behavior of components throughout the LC environment. By employing the Monte Carlo principle, the global system behavior can be simulated. They also give some solutions for risk mitigation according to the simulation. Vulnerability discovery plays a crucial role in mitigating potential risks within a system, and it has witnessed significant advancements with the progress of detection technologies. %\cite{luo2023vulhawk,cui2022vrust,wan2022too}
    Liu et al. \cite{liu2020cd} introduced a software vulnerability detection system, which leverages deep learning and domain adaptation techniques to address the challenge of software vulnerability detection. The system harnesses the automatic feature representation capabilities of deep learning and the domain adaptation framework to discover various vulnerabilities in heterogeneous projects and reduce software security risks. Wan et al. \cite{wan2022too} proposed a dynamic testing method for  semantic denial-of-service vulnerabilities, and accurately discovered nine unknown vulnerabilities in the planning of the actual open source autonomous driving system. Code reuse leads to the propagation of vulnerabilities. Luo et al. \cite{luo2023vulhawk} developed an intermediate representation function model to achieve cross-architecture binary code search through an entropy-based adapter and progressive search strategy, and tested on seven different tasks to prove the robustness of the model. Cui et al. \cite{cui2022vrust} proposed an automated vulnerability detection framework VRust for Solana smart contracts. The framework automatically detects potential vulnerabilities in contracts by analyzing the intermediate representation of translated Rust source code and program data flow. However, in real scenarios, network or system risk prediction faces more difficulties and challenges, and it is necessary to enhance the robustness of risk prediction in terms of consistency of execution, traceability of behavior, and integrity of record. These aspects can significantly improve the reliability and effectiveness of network or system risk prediction.

    Consistency of execution helps ensure smooth collaboration among multiple systems or entities involved in cross-domain risk detection. In real scenarios, achieving unified execution operations in cross-domain interactions over heterogeneous networks contributes to informed risk management and resource optimization. Ding et al. \cite{ding2020security} designed a resource management algorithm for heterogeneous integrated networks. During the process of collecting and managing heterogeneous resources in the heterogeneous network, the algorithm uses information security transmission technology to ensure the safe collection of resources and uses the improved management algorithm of heterogeneous resources to realize the security management of heterogeneous integrated network resources. Guo et al. \cite{guo2020master} introduced a reliable cross-domain authentication mechanism applied to the IoT. To achieve cross-domain authentication between heterogeneous IoT domains, the mechanism uses a master-slave blockchain architecture to ensure cross-domain privacy security. To achieve trusted authentication, an improved Byzantine fault-tolerant device based on the reputation value model (RIBFT) is used to conduct a credible assessment. Xuan et al. \cite{xuan2021cross} devised a certificate-less cross-domain authentication scheme that possesses the capability to distinguish parameters. The scheme relies on the principles of certificate-less public key cryptography and smart contract technology, which allows for the use of differentiated cryptographic system parameters for authentication between heterogeneous IoT networks, thus enhancing the security of cross-domain authentication for heterogeneous IoT networks. Hao et al. \cite{hao2022blockchain} proposed a lightweight architecture for consortium blockchain. The architecture utilizes a token accumulation mechanism for authentication of data access control and trust evaluation of requesting nodes. It supports cross-domain data sharing among Internet of Things users from different geographical locations. Li et al. \cite{li2022federated} introduced a trust mechanism hierarchy that relies on cooperative detection among blockchain nodes. Firstly, they employed federated learning to train a cross-domain unified behavior detection model which broke down data barriers and achieved cross-domain unified evaluation of device behavior trust. Based on this, they designed a layered trust mechanism based on federated detection combined with the transaction performance of blockchain. By dynamically evaluating devices based on behavior detection and blockchain transaction detection, graded trust management of devices is implemented. Sheff et al. \cite{sheff2016safe} aimed at the problem of safe scheduling in a federated environment and implemented a static detection compiler and a system that introduced a phased commit protocol to ensure the consistency and security of scheduling. %待补充
    Chen et al. \cite{chen2021xauth} proposed an efficient and privacy-preserving cross-domain authentication scheme, named XAuth, to address the cross-domain authentication issue in Public Key Infrastructure (PKI). The scheme utilizes Multiple Merkle Hash Trees to ensure the responsiveness of cross-domain data management and employs zero-knowledge proof algorithms to ensure privacy in cross-domain authentication. Lin et al. \cite{lin2022crossbehaauth} designed a time-aware cross-scenario keystroke dynamic authentication mechanism to address the issue of collecting a large amount of data every time the authentication scenario switches. The method improves the quality of data by selectively learning and encoding time information to achieve efficient behavior pattern transfer across scenarios. The method improves data diversity and cross-scenario applicability through a local Gaussian data augmentation method to enable consistent authentication across different scenarios. However, objectives such as cross-domain communication and heterogeneous integration still face several challenges. Cross-domain and heterogeneous networks typically entail distinct security policies and mechanisms. Achieving unified execution operations necessitates overcoming conflicts between diverse security requirements while ensuring that cross-domain operations do not introduce novel security vulnerabilities.

    Traceability of behavior plays a key role in risk detection. There is a great deal of uncertainty in the task of risk prediction. In instances of prediction failure, it is necessary to promptly trace the risk behavior to address vulnerabilities and minimize losses. Traceability of behavior is mainly categorized into two types: internal traceability and relation traceability. Internal traceability involves tracing user behavior within the network environment to distinguish potential risk behaviors or to trace the source of suspicious behavior that has already occurred. Relation traceability refers to analyzing a series of behaviors within the network environment, tracing suspicious associated behaviors, or tracing attacks from outside the network environment. Zhang et al. \cite{zhang2017towards} presented a secure s-health system designed for cloud service environments. The system introduces a decryption component that is integrated with the user's information during key retrieval. Once integrated, the component remains fixed and prevents key owners from re-randomizing, thereby establishing a binding between the user's information and enabling the tracking of a series of user behaviors. Lin et al. \cite{lin2021btdetect} proposed a method for detecting internal threats in cloud environments based on behavior traceability. First, the method analyzes the call rules of the cloud service interface to construct the complete behavior process of the call. Then, it uses the behavior tree construction algorithm to generate a legitimate behavior tree describing the behavior of cloud users. Subsequently, behavior trace points are set up to capture call behavior information on the service invocation interface of cloud services. Finally, user interface call information is matched with legitimate behavior trees through keyword matching to trace the source of malicious user behavior. Yu et al. \cite{yu2021blockchain} proposed a blockchain-enhanced security access control scheme to support traceability and revocability in IoT. Specifically, the scheme involves blockchain-based authentication to store all user information and public keys. Subsequently, system parameters are issued by administrators to users, along with a unique  parameter embedded in private keys. The scheme enables tracing malicious behavior by utilizing the parameter in private keys and revoking malicious users accordingly. Wang et al. \cite{wang2023composite} proposed a method for tracing associated events using a composite blockchain structure. Firstly, a storage structure model for the composite blockchain was constructed to achieve data association storage. Secondly, by obtaining the source entity block, an event association graph was constructed by using a source tracing method based on the Apriori algorithm. Finally, the entities were subjected to risk assessment using reinforcement learning. Zhu et al. \cite{zhu2022attacker} proposed an Ethereum attack traceability method based on graph analysis. They applied graph analysis techniques to analyze the behavioral characteristics of attackers and the relationships among them. Additionally, they used RPC mechanisms to trace the related attackers and attack sources. Behavioral tracing also encounters several challenges. More sophisticated attack behaviors excel at concealing their attack features, rendering themselves indistinguishable from normal behavior patterns. As a result, these behaviors can lurk within the system for an extended duration, which is currently difficult to trace.

\begin{table}[t]\scriptsize
   \centering
   \renewcommand{\arraystretch}{1.7}
        \caption{Summary of  behavioral benignity authentication}
          \resizebox{1\textwidth}{!}{
                \begin{tabular}{m{2.7cm}<{\centering}m{3.8cm} m{4cm} m{3.8cm} }
        \toprule[\heavyrulewidth] % Top line with heavier thickness
    \multicolumn{1}{c}{ \textbf{Object} }&\multicolumn{1}{c}{\textbf{Description}}& \multicolumn{1}{c}{\textbf{Characteristics}}& \multicolumn{1}{c}{\textbf{Issues}}\\
%        \begin{tabular}{p{3.1cm} p{3.8cm} p{4.1cm} p{3.8cm} }
%  %\begin{tabular}{>{\centering\arraybackslash}p{3.2cm}>{\centering\arraybackslash}p{4.5cm}>{\centering\arraybackslash}p{2.6cm}>{\centering\arraybackslash}p{2.6cm}}
%        \toprule[\heavyrulewidth] % Top line with heavier thickness
%        \textbf{Object} & \textbf{Description} & \textbf{Characteristic} & \textbf{Issues} \\
        \midrule
       Predictability of risk
       \cite{hu2018security,wang2023research,LiQiao2014,yang2022network,ghazel2009using,liu2020cd,wan2022too,luo2023vulhawk,cui2022vrust}  & Predictability of risk refers to the distinction of potential behavioral risks in the network environment. & (1) Prevention of potential attacks. (2) Remediation of exploitable vulnerabilities.  (3) Reduction of losses through early warning. & Due to the inherent uncertainty of evolving threats, ensuring the robustness of risk prediction needs to be complemented with other security technologies.
    %   It is necessary to enhance the robustness of risk prediction in terms of consistency of execution, traceability of behavior, and integrity of record.
       \\
        \midrule % Line to separate the Goal section
    Consistency of execution
    \cite{ding2020security,guo2020master,xuan2021cross,hao2022blockchain,li2022federated,sheff2016safe,chen2021xauth,lin2022crossbehaauth}  &Consistency of execution primarily refers to the method of credibility evaluation and potential risk detection that can perform consistent operations across complex and diverse networks. & Ensuring the seamless execution of operations across various domains and heterogeneous networks in multiple scenarios facilitates unified risk management and resource optimization. & Achieving consistency of execution requires compatibility with different resource forms and operating mechanisms in cross-domain interactions over heterogeneous networks. \\
        \midrule % Line to separate the rest of the rows
     Traceability of behavior
     \cite{zhang2017towards,lin2021btdetect,yu2021blockchain,wang2023composite,zhu2022attacker} & Traceability of behavior refers to the traceability of potential risks or suspicious behaviors in the network environment. &(1) Tracing suspicious behavior before it becomes a threat. (2) Tracing malicious behavior back to its source and close loopholes. (3) Tracing the source of related behaviors to reveal hidden attack chains.  & More sophisticated attackers excel at hiding their attack signatures and making attack behaviors indistinguishable from normal behaviors so that they lurk in systems for extended periods of time to carry out continuous attacks.\\
        \midrule % Line to separate the rest of the rows
   {\textbf{}} Integrity of record
   \cite{li2022blockchain,javaid2018blockpro,patil2020efficient,barbareschi2019puf,wei2020blockchain} & Integrity of record means that user data in a trusted network environment should remain complete, accurate, and non-tamperable. & (1) Improvement of risk prediction accuracy through complete and accurate data. (2) Enhancement of traceability analysis reliability. &  Protecting record integrity faces multiple data tampering attacks and introduces performance overhead and system complexity like data protection in distributed systems.\\
        \bottomrule[\heavyrulewidth] % Bottom line with heavier thickness
    \end{tabular}}
    \label{sumbba}
\end{table}

    Integrity of record can provide a reliability and accuracy guarantee for risk prediction and behavior traceability. Integrity of record includes integrity, accuracy, and tamper resistance of data. Current research is mainly focused on data integrity protection and tamper resistance. Li et al. \cite{li2022blockchain} integrated the trusted execution environment SGX with blockchain technology to construct a privacy-preserving multimedia authentication system. The system utilizes SGX to create a trusted execution environment and employs PhotoChain's hybrid storage mode to only store the hash values of the photos on the blockchain. By using blockchain to ensure data integrity, the system does not add a storage burden to the blockchain. Javaid et al. \cite{javaid2018blockpro} proposed a solution based on Physically Unclonable Functions (PUFs) and blockchain, known as BlockPro, for enhancing the security of data sources and ensuring data integrity in IoT environments. Specifically, the characteristics of PUFs can be used to establish a data source to ensure a unique source, and the data storage method of Ethereum ensures data integrity. Patil et al. \cite{patil2020efficient} proposed an efficient privacy-protecting authentication protocol that combines blockchain technology and PUFs. The protocol employs a decentralized digital ledger using blockchain smart contracts to resist attacks from data tampering, thereby ensuring security in IoT environment. Additionally, by integrating the uniqueness and tamper-proof properties of PUFs with blockchain, the protocol ensures unique device IDs and data integrity in the IoT. Barbareschi et al. \cite{barbareschi2019puf} proposed a mutual authentication scheme relying on the use of PUFs. The scheme employs PUFs' characteristics of being unclonable, unique, and tamper-evident to protect edge nodes in IoT from physical attacks and data tampering. Wei et al. \cite{wei2020blockchain} proposed a solution to address data security concerns in cloud computing by integrating blockchain technology. The approach involves deploying a distributed virtual machine proxy model in the cloud through the use of mobile agents, which ensures reliable data storage. Furthermore, the solution leverages a blockchain-based data integrity protection framework and generates hash values for corresponding files using the Merkle hash tree. Data alterations are then monitored via smart contracts and alerts are triggered in case of any tampering. In the process of data integrity protection for cross-domain transmission, researchers need to find a balance between security and efficiency, so as to provide more reliable support for risk prediction and behavior traceability.

     Based on the aforementioned analysis, characteristics and issues of different objects of behavioral benignity authentication are summarized in Table \ref{sumbba}.

\section{Challenges and future research directions}\label{challenge}
%\vspace{-0.1in}
%-----------------------------------------

\begin{figure}[t]
  \centering
  \includegraphics[width=0.7 \textwidth] {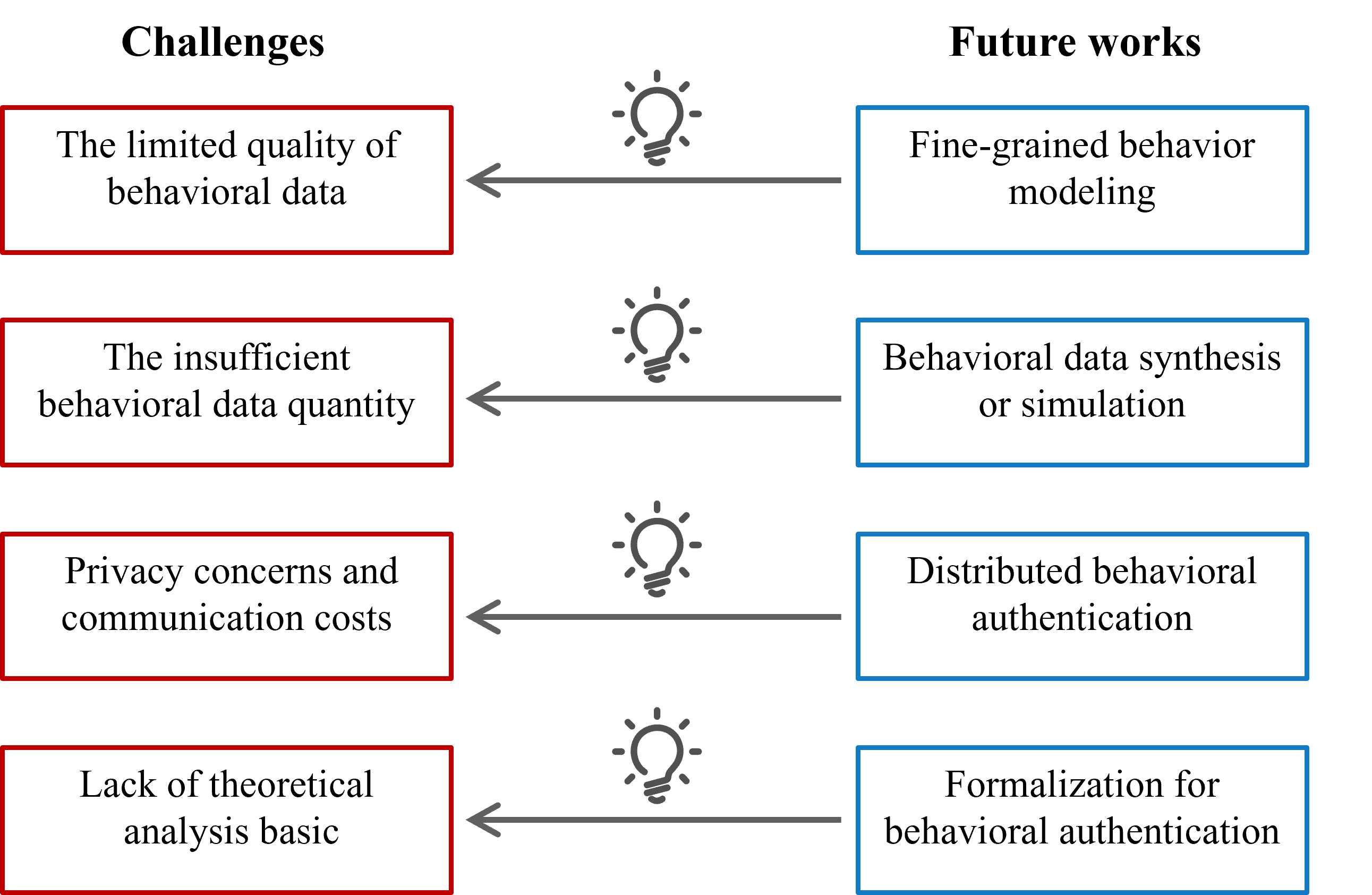}
% \vspace{-0.18in}
   \caption{Overview of main challenges and future research directions for behavioral authentication.
   }
  \label{figure:challenge}
    \vspace{0.1in}
\end{figure}

We thoroughly examine the main limitations associated with existing behavioral authentication methods and discuss innovative research directions that have the potential to significantly augment the applicability and effectiveness of behavioral authentication. A comprehensive overview of limitations and future research directions is shown in Figure \ref{figure:challenge}.
\subsection{Challenges of behavioral authentication}
%.We analyze main limitations of the existing behavioral authentication, and in light of these limitations, we envisioned future research directions for behavioral authentication. This aims to further enhance the application scenarios and potential of behavioral authentication. The overview of
%我们分析了现有行为认证存在的一些局限性，同时针这些局限性，展望了未来的行为认证的研究方向，以进一步提升行为认证的应用场景和潜力，总体的关联如图所示
\subsubsection{The limited quality of behavioral data}
%Behavior modeling data is typically of limited quality. This data limitation can generally be attributed to three factors: data collection and processing, privacy protection, and business attributes. In terms of data collection and processing, the inherent difficulty in collecting or processing the data can result in low data quality . Regarding privacy protection, data quality is restricted by the privacy requirements of users or service providers . In terms of business attributes, the inherent characteristics of the business make it challenging to model user behavior effectively . For instance, in the context of online payment scenarios, the extremely imbalanced ratio of black and white samples of account fraud poses a significant difficulty in behavior modeling.
%Designing feasible data augmentation methods is a crucial prerequisite for achieving effective behavioral authentication, aiming to establish high-quality behavior models using low-quality behavioral data.

How to overcome the limitations of behavioral data quality in the modeling process remains a primary concern. These limitations can be generally attributed to three factors: behavioral data collection and processing, privacy protection, and business attributes. In terms of data collection and processing, inherent difficulties in behavioral data collection or processing can result in low data quality \cite{jing2019thinking}.
%For example, during the collection of behavioral data, issues such as duplicate or missing data may arise, leading to inaccurate or even erroneous results in behavioral data analysis. Therefore, rigorous data cleaning and validation are essential before behavioral data anaylsis.
  For example, during the collection of behavioral data, issues such as duplicate or missing data may arise due to unreliability in collection equipment or inconsistencies in data sources, leading to inaccurate or even erroneous results in behavioral data analysis. These undesirable results have a direct impact on the quality and effectiveness of behavior modeling. Therefore, to guarantee the high quality of behavior modeling, it is imperative to have a rigorous process of data cleaning and validation.
In terms of business attributes, the inherent characteristics of commercial activities pose challenges in effectively modeling user behavior \cite{bauder2018effects, kaur2018comparing}.
In scenarios such as online payment services, the highly imbalanced ratio of fraudulent accounts to legitimate accounts presents significant difficulties in behavior modeling.  For interaction data across diverse terminal devices, the distinction between normal and abnormal data is frequently ambiguous. Relying solely on attribute data has limited effectiveness.
Designing feasible data augmentation methods is a key prerequisite for achieving effective behavioral authentication and aims to establish high-quality behavior models using low-quality behavioral data.

\subsubsection{The insufficient behavioral data quantity}
%The challenge posed by the insufficient quantity of behavioral data lies in its detrimental impact on the effectiveness of behavioral authentication systems. With limited data available for training, it becomes difficult to create robust machine learning models capable of accurately distinguishing between legitimate and malicious behaviors. This scarcity also increases the risk of biased or skewed models, as they may learn from a narrow subset of behaviors, leading to decreased reliability in authentication performance. Furthermore, the lack of adequate data inhibits the generalizability of authentication systems, as models may struggle to adapt to new or evolving threats without comprehensive training data. Addressing the challenge of insufficient behavioral data quantity requires exploration and development of approaches to alleviate the insufficient behavioral data quantity.
Behavioral authentication is also data-driven, and the insufficient behavioral data quantity  may influence the effectiveness of behavioral authentication systems.
%Usually, in anomaly detection scenarios, there exists following issues of data: rarity of anomalies, label imbalance, and sampling bias.
With limited behavioral data available for training, it becomes difficult to create robust behavioral authentication models capable of accurately distinguishing between legitimate and malicious behaviors. This scarcity also increases the risk of biased or  unfair models, as they may learn from a narrow subset of behaviors, leading to decreased reliability in authentication performance. Furthermore, the attackers may not repeat previously detected or blocked methods to attack the system,
which also intensifies the risks associated with the system \cite{eisenberg2001systemic}. The lack of behavioral  data inhibits the generalizability of authentication systems, as models may struggle to adapt to new or evolving threats without comprehensive training data. Therefore, addressing the challenge of insufficient behavioral data quantity is crucial for the development of  highly reliable behavioral authentication systems.
\subsubsection{ Privacy concerns and communication costs}
Under the stricter privacy protection regulatory constraints, such as CCPA,
GPDR, and DSL \cite{stallings2020handling,godinho2021consumer,chen2021understanding}, behavioral data are more tightly controlled.  People are also increasingly valuing privacy as they become more aware of the risks associated with data breaches and unauthorized surveillance. Heightened concerns about personal information security have motivated individuals to demand greater transparency and control over how their data is collected and used.
These changes directly lead to the difficulty of implementing existing centralized authentication methods for collecting behavioral data from different agents.
%这些因素直接导致现有集中式收集不同主体行为数据的认证方法难以实现。
%Privacy protection also has a significant impact on data quality, as it affects the types of behavioral data that can be collected, within the privacy constraints of users and service providers \cite{stallings2020handling, godinho2022consumer, cai2022demystifying}.
%Behavioral data is classified into general behavioral data, important behavioral data, and core privacy data, with different levels of protection measures applied. Strict protection is enforced for core data, while other behavioral data is subject to varying access requirements based on regulations. However, in practical application scenarios, accurately assessing the privacy of behavioral data can be challenging due to factors such as business interests and insecure network connections.
Despite the advancements in hardware technology and the development of $5$G technology, the current computational resources and latency in behavioral authentication also face significant challenges.
%In terms of computational resources, it is difficult to adequately handle highly heterogeneous behavioral data \cite{heckman2001micro}.  这直接导致承重的通信开销
Behavioral data exists in various forms such as text, voice, and video, and the required models for training are of high complexity \cite{phan2020covernet}. This leads to excessive consumption of computational resources and  significant communication costs.
For example, when there is a change in the existing behavioral authentication patterns, it becomes necessary to update the entire model, which not only affects the usability of authentication but also results in unnecessary waste of computational resources. It is of great significance to determine models that are compatible with resource constraints and establish efficient knowledge transfer mechanism of multimodal behavioral data.
Regarding authentication latency, current mainstream authentication schemes still rely on the transmission of frequent behavioral data packets to achieve centralized behavior modeling. However, there is distance between different terminal devices, and centralized collection takes time to transmit behavioral data. Even a delay of a few milliseconds, while browsing the internet or attempting to connect to a smart refrigerator, may only affect user experience.
Additionally, the centralized authentication relies on a single centralized entity, leading to excessive concentration of authority, and single point failures can result in severe system issues.
However, in scenarios such as automated remote surgery or autonomous driving \cite{derman2018continuous,gupta2019tactile}, a few milliseconds of latency could lead to fatal accidents. Therefore, establishing more flexible deployment schemes for behavioral authentication not only ensures lower latency but also avoids unnecessary network congestion and reduces unnecessary bandwidth costs.
%此外，分中心化的认证依赖唯一的中心化机构，权限过于集中，存在单点故障等问题 。
%Despite the remarkable progress in hardware technology and the emergence of 5G, which has significantly enhanced network transmission capacity, there are still notable challenges in the speed of behavioral authentication.  One of the primary reasons for this limitation is the heterogeneity of behavioral data \cite{heckman2001micro}, which can encompass various forms such as text, voice, video, and more.
%The diverse nature of behavioral data introduces complexities in training the models necessary for behavioral authentication.  Developing accurate and robust models that can effectively process and analyze different types of behavioral data requires substantial computational resources and sophisticated algorithms.  As a result, the speed of behavioral authentication is constrained by the computational complexity associated with training and deploying these models.
%Moreover, the issue becomes even more pronounced in latency-sensitive authentication services and scenarios.  For applications where real-time authentication is critical, such as financial transactions or access control systems, the limitations in speed can have a significant impact on user experience and system performance.
%尽管业内对于人工智能领域的诸多问题尚未达成明确的共识，但行业对大模型的发展认知正在逐渐清晰，其中一点就是数据质量和数据量将是下一阶段大模型能力涌现的关键。

%数据量将是下一阶段行为认证的关键
\subsubsection{Lack of theoretical analysis basic}
The existing foundational theoretical efforts  primarily focus on conducting specialized theoretical analysis from specific perspectives, considering the constrained conditions of user behavioral data.
%However, the foundational and generalizable nature of these theoretical analysis remains limited.
On one hand, these studies excessively rely on the employed models, making it hard to perform prior evaluation and analysis of data utility. This circumstance prevents accurate anticipation of model performance before data utilization and hinders the determination of whether the chosen models possess sufficient applicability for particular tasks or applications. On the other hand, the data-driven paradigm is difficult to cover or traverse all possible scenarios, consequently impeding the attainability of a general applicable theoretical framework.
The diverse requirements arising from different data sources and tasks may necessitate distinct methods and models for conducting behavioral authentication. In summary, the existing relevant research lacks theoretical analysis basic of behavioral authentication, which makes it difficult to evaluate the performance of behavioral authentication in practical applications and hampers the provision of theoretical guidance for optimizing approaches. Hence, the exploration of the formalization for behavioral authentication holds significant significance, akin to the guiding role of information theory's fundamental limits in digital communication technology. It will not only offer architectural guidance for high-performance models and algorithm design in behavioral authentication but will also provide systematic metrics for evaluating specific  performances. Simultaneously, it will provide a theoretical foundation and analytical methodologies for understanding the mechanisms of behavior modeling in typical services, holding particular importance for estimating data utility in the data handover phase.
Efforts towards addressing this issue contribute to a better comprehension of essential issues and drive the advancement and progress of the entire field.
%现有的一些基础理论方面工作，主要针对用户行为数据的受限条件，开展特定角度的相关理论分析，但是这些理论分析的基础性与通用性不强。具体来说，一方面，这些研究，过于依赖所采用的模型，难以做到数据效用事前的评估与分析。这种情况下，导致无法在开始使用数据之前准确地预测模型的性能如何，也无法确定所选择模型是否在特定任务或应用中具有足够的适用性。另一方面，数据驱动的范式所有可能的情况难以覆盖或遍历，也导致了通用性的理论框架难以企及。不同的数据源、不同的任务都可能需要不同的方法和模型来进行行为认证，而且这些需求可能在不同的情境下发生变化。总而言之，现有相关研究缺乏通用性的行为认证的形式化，难以对实践应用中的行为认证性能评估提供泛化性的理论依据，以便对方法的优化方向提供理论指导。因此建立基础性的行为理论形式化方法与分析框架，为通用性数据效用事前分析机制的基础理论提供范例性研究具有重要意义。这方面研究的重要意义类同于信息论基本极限理论对于数字通信技术的指导意义，将为行为认证的高性能模型和算法设计提供架构性指导，为具体模型性能评价提供系统化指标；同时为典型网络服务行为建模机理提供理论基础和分析方法，对于数据交接环节的数据效用事前估计具有重要意义。尽管通用性的行为认证的形式化很难达到，但是其有助于更好地理解和优化行为认证领域的关键问题，从而推动整个领域的发展和进步。

\subsection{Future research directions of behavioral authentication}
\subsubsection{Fine-grained behavior modeling}
Future research should focus on developing the behavior modeling method that fully utilizes limited data, taking into account the interaction and collaboration between behaviors from a fine-grained perspective. Fine-grained behavior modeling leverages deep and rich information from the behavior data, which can provide accurate and reliable behavior distribution for behavioral identity, conformity, and benignity authentication.

We are committed to proposing a fine-grained behavior modeling framework to enhance the limited behavior data. It considers the associations of behavior across multiple dimensions such as behavior sequences, subject correlations, and attribute distributions from complex and diverse behavioral data in both virtual and real spaces. Through holistic learning of abundant behavioral information, we can better understand the underlying semantic meanings in behavioral data, and subsequently apply the learned semantics for enhancing the authentication performance.

\begin{figure}[t]
  \centering
  \includegraphics[width=0.99 \textwidth] {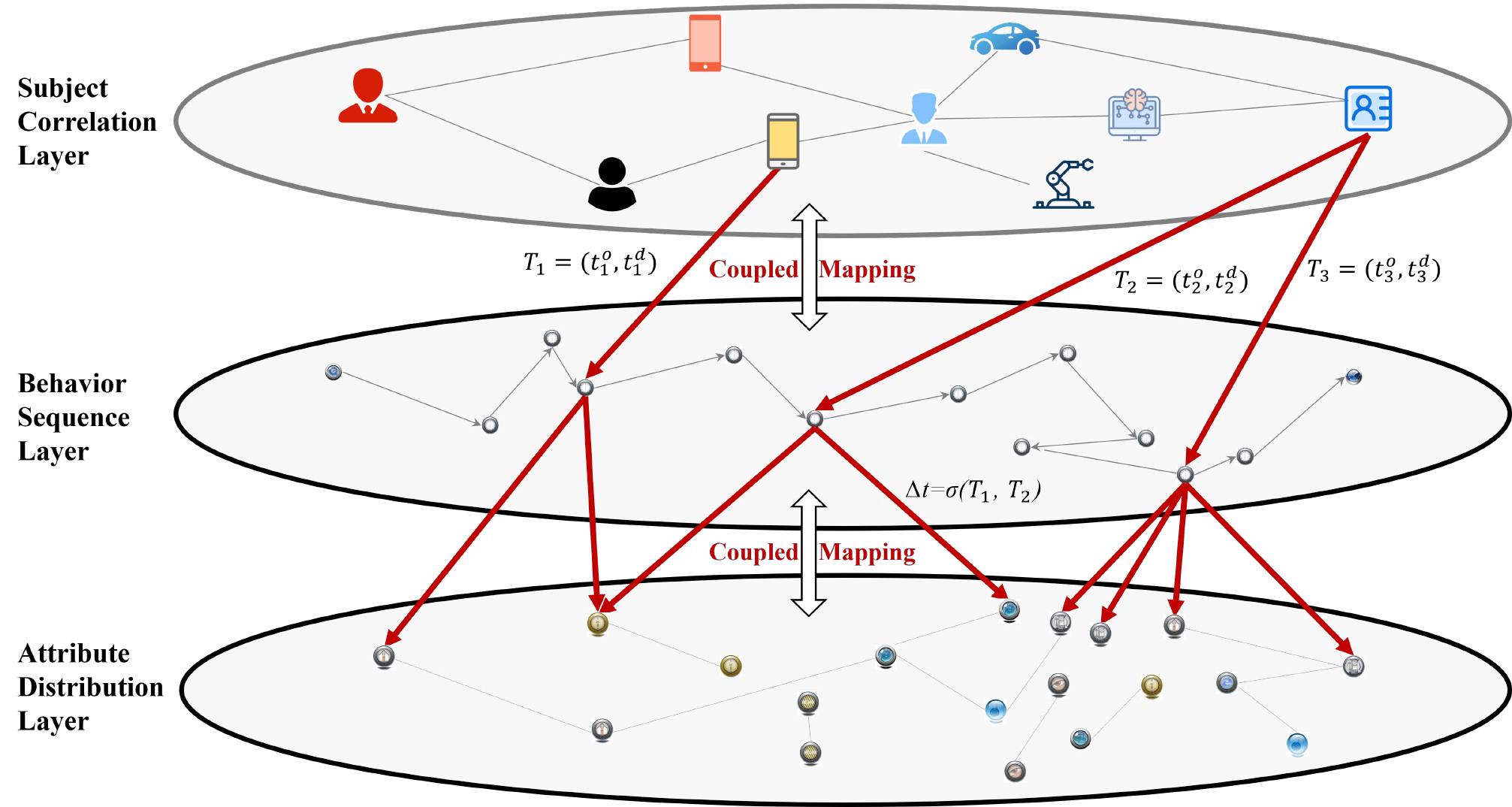}
% \vspace{-0.18in}
   \caption{The fine-grained behavior modeling framework that consists of the subject correlation layer, behavior sequence layer, and attribute distribution layer.  The subject correlation layer reflects the association between subjects and is coupled with a mapping relationship to the behavior sequence layer. There is a time-annotated edge with time information $T=(t^o,t^d)$ between each subject and behavior, where time $T$ indicates the starting time $t^o$ and ending time $t^d$ when the subject initiates the behavior.
  In the behavior sequence layer, different behaviors occur in a clear sequence, and we denote $\Delta t=\sigma(T_1,T_2)$ as the time difference between two behaviors $T_1$ and $T_2$. A behavior is described in detail by several attributes, and their associations are reflected as the coupling mapping between the behavior sequence layer and the attribute distribution layer.}
  \label{figure:integration}
%  \vspace{-0.1in}
\end{figure}
Correspondingly, as shown in Figure \ref{figure:integration}, a potential solution is to establish a three-layer knowledge graph structure, including the subject correlation layer, behavior sequence layer, and attribute distribution layer.
In the subject correlation layer, subject correlation associations are modeled by introducing the similarity between subjects and their social relevancies. Specifically, the behavior sequence layer and attribute distribution layer provide characterization of the latent behavior space encompassing all behaviors, allowing generalization of subjects from instantiated modeling objects in network services to any behavioral attributes (such as population, individual and class-based subjects). Multi-dimensional intelligent synthesis strategies are designed to cooperatively generalize the distribution of subject models, thereby mitigating the technical bottleneck of insufficient data in security and safety authentication.
In the behavior sequence layer,  the temporal relationships of behaviors, such as the order in which behaviors occur, are modeled. We aim to propose a variable-length sequences modeling solution that  specifically can be divided into prefix, infix, and suffix partitions based on behavioral order, and adaptive partitions based on behavioral windows. Compared to traditional sequence modeling techniques, it combines behavior intent recognition and time dependency learning. By extracting information from behavioral context and intent, it alleviates the inefficiency issues of authentication techniques on limited data.
In the attribute distribution layer,  co-occurrence associations between attributes from the same behavior are extracted. For example, multiple attributes that appear together in a behavior can serve as co-occurrence relationships between attributes.
We conceptualize behavior as a stable system composed of internal associations at the topological level. By designing customized behavioral scanners, we quantify the associations between fine-grained behavioral attributes as measurable graph objects. It excludes confounding information arising from knowledge-driven interference within behaviors, focusing instead on potentially internal associations between fine-grained behavioral attributes as driven by the data. So it can eliminate behavioral noise caused by erroneous or outdated human annotation.
Based on the three customized layers,  two types of coupled mapping, i.e., subject behavior mapping and behavior attribute mapping, are devised. The former realizes the effect of subject collaborative filtering through the interaction between subjects and behaviors. Subjects with similar preferences are reflected in the subject correlation layer. The correlations between subjects are further described by the interaction with time between subjects and behaviors.
The latter realizes the effect of content collaborative filtering through the interaction between behaviors and attributes, which reflects behaviors with similar attributes in the behavior sequence layer.
Through hierarchical modeling of behavior, in addition to introducing more associations to enrich the description of behavior, we can perform customized behavioral authentication based on specific associations within each layer. The coupled mapping between layers further allows aggregating information from other layers into a single layer, thereby obtaining high-density behavioral semantics (information stacking and reduced information carriers jointly improve density) under limited behavioral data to support different authentication tasks.

\begin{figure}[t]
	\centering
		\includegraphics[width=1\columnwidth]{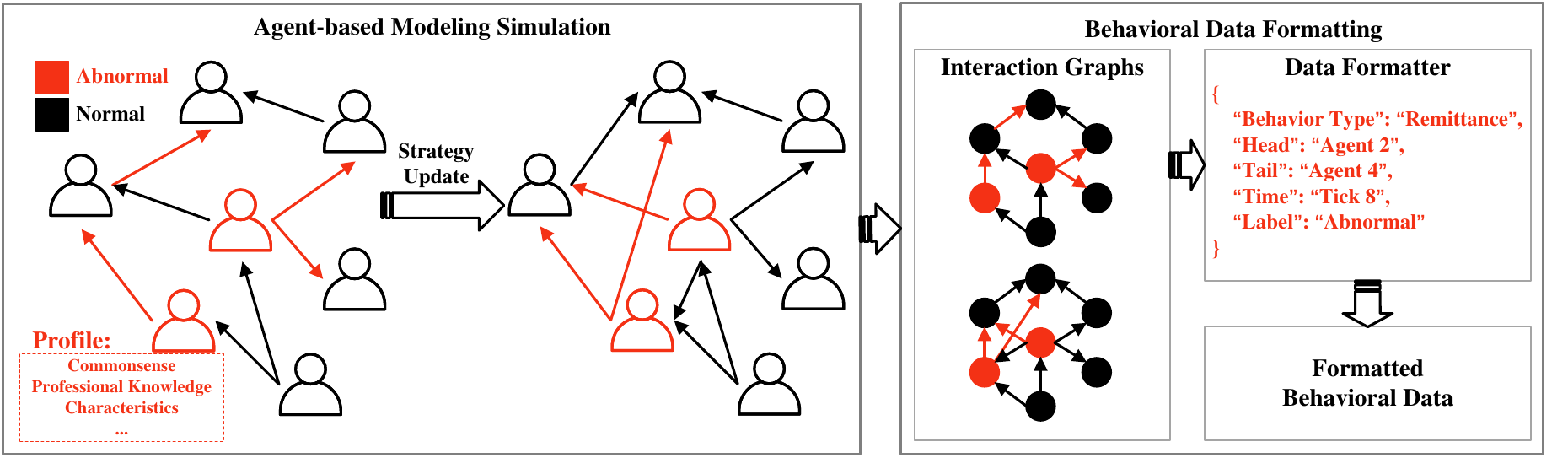}
	\caption{The illustration of behavioral data simulation. First, agents with open-domain reasoning ability and detailed profiles are launched in an ABMS system. Second, the interactions between agents are converted into interaction graphs. Third, these graphs are formatted into the required form through a data formatter. Finally, these data are regarded as simulated behavioral data to supplement the origin dataset.}
		\label{fig:DQ}
\end{figure}
\subsubsection{Behavioral data synthesis or simulation}
The challenge of insufficient behavioral data consists of two parts.   On the one hand, in scenarios with a defined process, behavioral data synthesis methods are required to supplement the datasets, especially samples of illegal behavior. Data synthesis methods can be utilized to expand existing datasets with data of the same distribution, and the gap between normal behavior samples and illegal behavior samples could be narrowed. On the other hand, in scenarios with complex interactive processes, the interactions of individuals lead to chaos so that there does not exist a fixed distribution of illegal behaviors. Therefore, synthesis methods creating behavioral data of the same distribution no longer work, and we suggest behavioral simulation as an effective resolution.

%For behavioral data synthesis, we suggest a novel GAN-based meta-learning anomaly detection framework, referred to as EM-GAN. EMGAN consists of a generator $G$ and two discriminators $D_1$ and $D_2$. The generator incorporates an autoencoder to map the data into a high-dimensional space, where the outlier characteristics of anomalies are accentuated through a specially designed mapping function. The interplay between $G$ and $D_1$ aims to embed data resembling the real data distribution, while the interaction between $G$ and $D_2$ enhances the separation between positive and negative sample distributions. The adversarial training between $D_1$ and $D_2$ further enhances the distinction of anomalies.
%对于行为数据合成，现有一些GAN等优秀方法/cite{}，但是他们哪方面存在不足（The generator incorporates an autoencoder to map the data into a high-dimensional space, where the outlier characteristics of anomalies are accentuated through a specially designed mapping function. 离群点难以捕获吗？？？），未来可以考虑结合原学习等结合行为正负样本之间的距离（between $G$ and $D_2$ enhances the separation between positive and negative sample distributions）。
For the synthesis of behavioral data, the existing methods exhibit sensitivity to outliers, which may  compromise the quality of generated samples in the latent space.
% Additionally, handling discrete data poses a relative challenge for existing techniques.
A potential approach to address these limitations is to combine a generator and two discriminators. The generator incorporates an autoencoder to map the data into a high-dimensional space, where the outlier characteristics of abnormal behavior are accentuated through a specially designed mapping function. The interaction between the generator and the first discriminator enhances the similarity between the generated data distribution and the real data distribution, while the interaction between the generator and the second discriminator enhances the separation between
the distributions of normal and illegal behaviors. The adversarial training between the first  discriminator and the second discriminator further enhances the distinction of illegal behaviors.
Simulation technologies have developed for many decades, consistently regarded as a powerful tool for analyzing complex systems. Agent-based modeling simulation (ABMS) represents an advanced simulation approach, characterized by the generation of individual-level behaviors for both items and agents \cite{smith2007agent}. In the era of data science, deep learning models have been harnessed to enhance the authenticity of individual behavior generation. This paradigm falls short in addressing the challenge of dynamic distributions, as agents empowered by deep learning models remain constrained by historical data and limited behavior sets. To enhance ABMS for simulating diverse behavioral data, we suggest a novel simulation framework centered around agents endowed with reasoning and learning capabilities. Notably, large language models (LLMs) have demonstrated open-domain reasoning abilities as agents. Leveraging these agents for simulating security scenarios (as depicted in Figure \ref{fig:DQ}), we provide them with security-related knowledge and detailed profiles encompassing motivation, behavior preferences, and other relevant factors, using a variety of models. Through the combination of foundational models with smaller-scale models, each agent is effectively characterized to respond thoughtfully to situational cues. Furthermore, these agents possess learning abilities, enabling them to devise novel strategies beyond historical data, similar to real-world attackers and defenders. In summary, an ABMS built upon this new type of agents equipped with hierarchical knowledge structures and adaptive learning mechanisms, offers a promising resolution for simulating behavioral data in complex scenarios.

\begin{figure}[t]
  \centering
  \includegraphics[width=1 \textwidth] {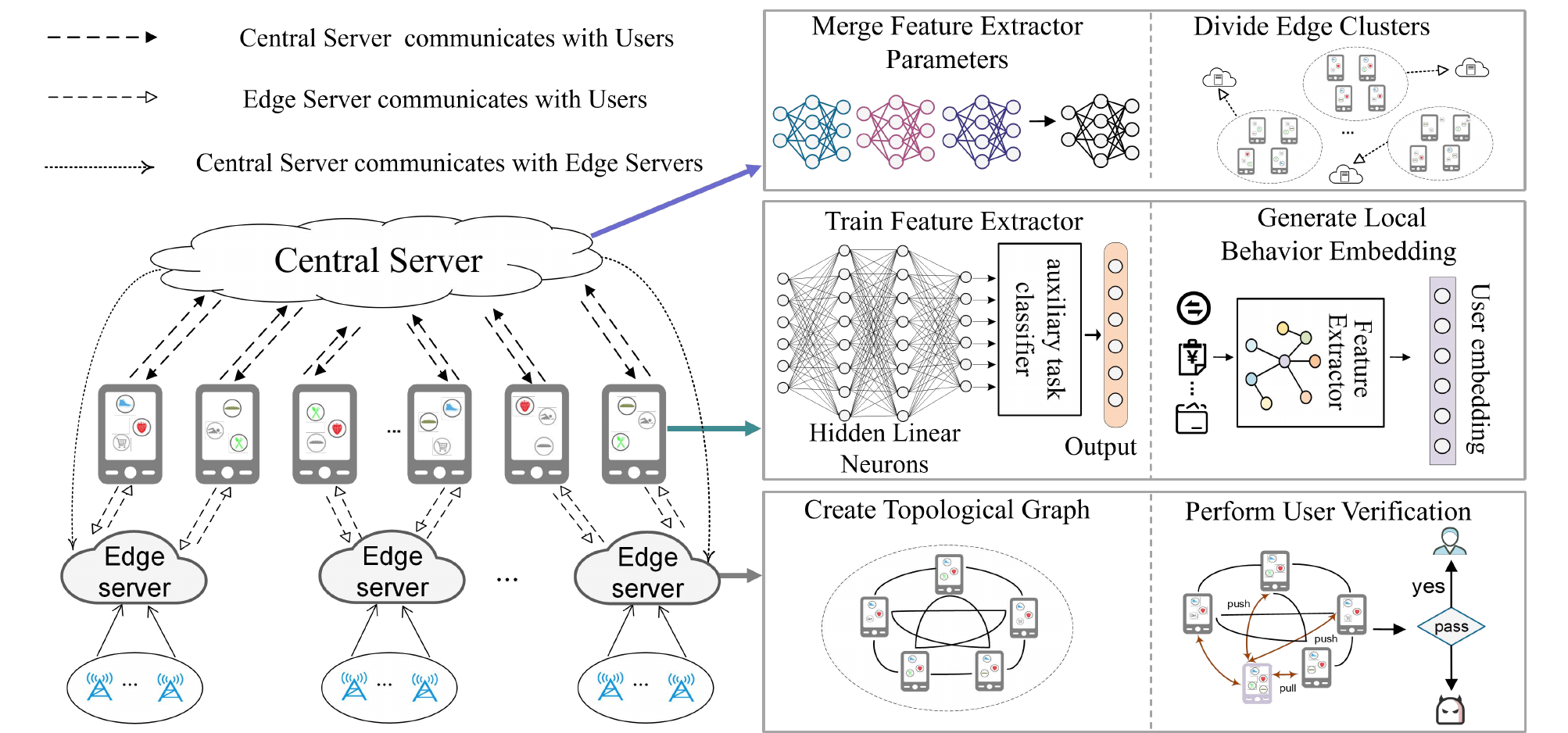}
% \vspace{-0.18in}
   \caption{The overall architecture of distributed behavioral authentication. The central server is responsible for collaborating with the terminals to construct feature extractors and coordinating the unified scheduling of edge servers. Edge servers are responsible for deploying specific authentication credentials and partitioning user clusters, as well as responding to specific authentication requests. Terminals are responsible for storing private sensitive data and initiating relevant requests during the verification process.}
  \label{figure:dis}
%  \vspace{-0.1in}
\end{figure}
\subsubsection{Distributed behavioral authentication}
In general, the effectiveness of behavioral authentication methods often relies on centralized data collection to construct classification models. However, with the increasing emphasis on stricter privacy protection regulations and the need for local data processing in distributed environment, the control and transmission of behavioral data become more challenging. It is necessary to develop distributed behavioral authentication models across different devices while minimizing the costs associated with transmitting privacy-sensitive behavioral data between devices.
The concept of computing power network has been implemented as the infrastructure continues to improve, enabling behavioral data to transition from single central deployment to diverse distributed deployment. This shift from centralized scheduling to distributed collaborative scheduling has provided a favorable opportunity for the advancement of behavioral authentication in near-real-time processing.
Sensitive behavioral data can be processed and stored locally, while other behavioral data can be decoupled, allowing for selective interaction with servers in different physical locations. Even if the parties involved in data sharing do not have identical definitions of sensitive behavioral information, negotiations can be conducted to determine the optimal approach for decoupling the sensitivity levels of behavioral information.
Distributed behavioral authentication systems can bring enhanced processing capabilities and reduce response time, which improves effectiveness and reliability of the authentication process significantly.

Therefore, as an essential computational paradigm, distributed learning holds great potential in designing viable architectures for behavioral authentication to achieve a balance among authentication performance, user privacy, and communication costs. An objective is to improve cloud-edge-terminal collaboration to fully leverage the centralized and intensive deployment capabilities of the central cloud, the low latency and high flexibility of edge servers, and the local processing power of the terminal devices.  The collaboration approach enhances the safety of behavioral data interactions. Through the collaboration of cloud computing, edge verification, and local data processing, it reduces the processing burden on the cloud center, lowers bandwidth load and authentication latency, and mitigates behavioral data transmission costs.

 Correspondingly, as shown in Figure \ref{figure:dis}, a distributed behavioral authentication framework that leverages cloud-edge-device collaboration  has been established,  further expanding the application scenarios of behavioral authentication. Specifically, the central server and devices work together to construct a feature extractor. To achieve this, an auxiliary task is defined to obtain behavioral embedding vectors based on a deep neural network. During this process, terminal data remains to be stored locally without the risk of privacy leakage.
 The local devices share the partial model parameters with the central server to ensure the measurability of different behavioral embedding vectors.
If the learning rate of the device is $l$ when the global parameters are updated in the $t$ round,
each device will perform parameters updates locally through
${w_{t+1}}^k = {w_t}^g - l\nabla{g_k}.$
The central server uses the operation of
${w_{t+1}}^{g} =\sum\nolimits_{k =1}^{K}\frac{n_k}{\sum_{k=1}^{K}{n_k}}\cdot{w_{t+1}}^{k}$
to aggregate parameters,
and returns  ${w_{t+1}}^{g}$ to different devices. The device updates the local model with the
new parameters for the next training round.
%Let $l({x_i},{y_i};{w})$ denote the loss of prediction on $(x_i, y_i)$ made with local parameters $w$.
%The objective function of the auxiliary task is defined as follows:
%${\min}~L\left({x_i},{y_i};{w} \right)=\frac{1}{n_k}\sum\nolimits_{i =1}^{n}l\left(x_i^k,y_i^k;w\right).$
 Once the feature extractor has been built and deployed
to various devices, corresponding behavioral vectors are generated locally by different devices, and then different devices send these vectors to the central server. When the central server receive behavioral representation vectors from various terminals, it calculates the similarity between these vectors. Subsequently, the authentication service is deployed to the optimal edge locations based on demand. For the deployment of edge servers, the optimization objective is centered around minimizing latency by:
%  $L = \min \limits_{r_j\in R}\sum_{s_k \in S}\sum_{a_i \in Z_k} q(a_i, s_k;r_j),$
$\min \mathcal{L} (r)= \sum_{s_k \in S}\sum_{a_i \in Z_k} q(a_i, s_k;r),$
 where $r$ represents one certain feasible edge deployment scheme,
% $L(r_j)$ denotes the total authentication latency,
  $Z_k$ represents all base stations covered by current authentication server $s_k$, $S$ represents all edge authentication servers, and $q(a_i, s_k)$ signifies the latency from base station $a_i$ to edge authentication server $s_k$. Furthermore, each edge server establishes a topological graph. For terminals with similar patterns, corresponding behavior profiles are created, and based on the results, a topological graph $G = (V, E)$ is constructed. The node set is denoted as $V = \{x | x \in X\}$, where $X$ represents the terminals within that cluster, and $E =\{(x, y) | x, y \in V\}$, where $(x, y)$ represents an edge between two terminals. The connection strength of credentials between different terminals is calculated and stored on the edge server. During the verification phase, the $k$-th terminal generates new behavioral data. For the privacy-sensitive data of this terminal, the input data is processed through the trained feature extractor to generate a vector representation, denoted as $\Phi_{k_i}$, which is then sent to the edge server. The final authentication result is returned by the edge server. This approach achieves distributed behavioral authentication by the collaboration of cloud, edge, and devices, that significantly reduces authentication latency while ensures authentication performance.
%The proposed framework represents only the initial effort towards enabling distributed behavioral authentication.
%This distributed behavioral authentication framework is also compatible with other local model and training strategies. For example, when dealing with highly heterogeneous and non-independent  datasets across different clients, adjustments can be made to the local model structure and training strategies within the framework. This involves training personalized local models through strategies like model decoupling and global model personalization \cite{tan2022towards}.

 The proposed framework  represents initial efforts towards enabling distributed behavioral authentication, which possesses good flexibility and compatibility. In scenarios where datasets from different clients are highly heterogeneous and non-independently distributed, local model structures and model training strategies can be replaced in our framework. Personalized feature extractors can be trained through techniques such as model decoupling and global model personalization \cite{tan2022towards,collins2021exploiting}. Additionally, the framework can be used in conjunction with many other technologies. When addressing privacy concerns, our framework can employ technologies such as differential privacy \cite{wei2020federated} to alleviate privacy issues. Simultaneously, it can utilize existing robust adversarial defense techniques \cite{chen2023practical,huang2021starfl} to defend against security threats during the behavior modeling process. Furthermore, when there is distrust among multiple clients, our framework can address this issue by introducing blockchain technology \cite{shayan2020biscotti,lu2019blockchain}. It is orthogonal to our framework, and thus can be incorporated into our framework for storing behavior profiles.
\subsubsection{Formalization  for behavioral authentication}
Inspired by the formal  hypotheses and principles in information theory, we analogize behavioral authentication to concurrent network communication since both the problems of behavioral authentication and concurrent network communication share the common goal of eliminating uncertainty. The fundamental problem in communication is for one end of communication to accurately or approximately reproduce the message selected by the other end.
%One feasible and generalized formal approach is to directly use formal methods based on behavioral attributes to express the measurement of signal strength. However, this approach might face challenges such as losing the semantic information of behaviors, overfitting the properties of fitted functions, and underestimating the expressiveness of behavioral authentication models. The theoretical results derived from this formal framework represent a suboptimal upper bound on data utility.

\begin{figure}[t]
  \centering
  \includegraphics[width=1 \textwidth] {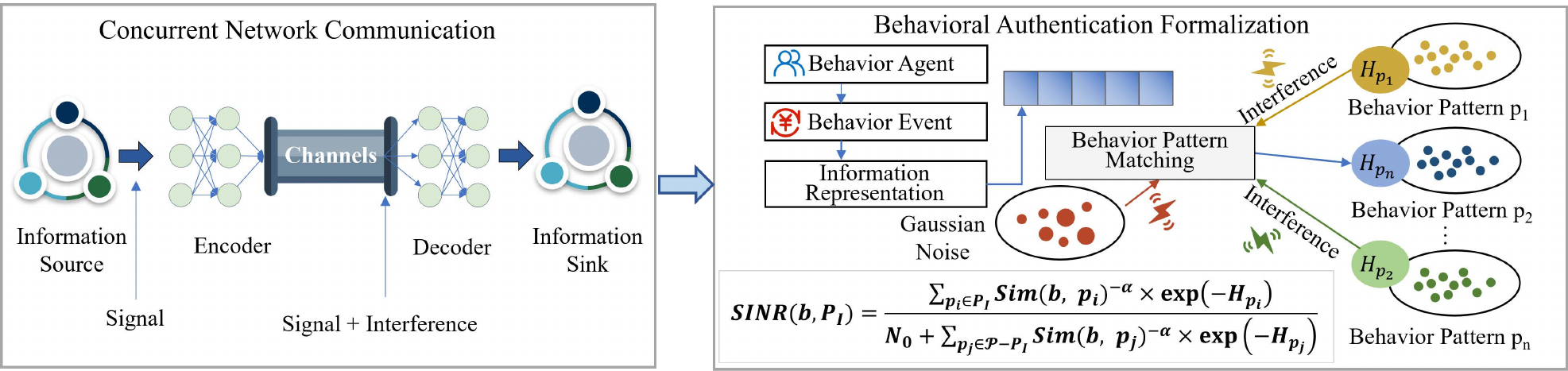}
% \vspace{-0.18in}
   \caption{The formalization for behavioral authentication. Inspired  by the formal hypotheses and principles in information theory, the ultimate essence of behavioral authentication is the elimination of behavioral uncertainty. This uncertainty can be quantified using a metric similar to the Signal-to-Interference-plus-Noise Ratio (\emph{SINR}) commonly found in communication.
   The impact on  behavioral authentication performance arises from interference by users with similar behavior patterns and the noise inherent in the data itself.
   }
  \label{figure:formula}
%  \vspace{-0.1in}
\end{figure}

We give the formalization method for behavioral authentication, which satisfies the formal hypotheses. As shown in Figure \ref{figure:formula},  we encode the complex features of data to obtain  behavior representations, which extract meaningful information and behavior event structures.
%Through deep learning networks, it extracts task-related information and data-related reconstruction structures.
%enhances the validity and reliability of the  .
% On the encoding end, the extraction module relies on the knowledge base and deep learning networks to extract semantic features from the raw data, combined with prior knowledge and data labels.
% The structural reconstruction module mines deep-hidden information in graph data using graph neural network models for graph data sources.
% The mapping module concatenates event information and structural information through a linear layer, reducing model complexity to prevent overfitting.
Ultimately, behavioral authentication aims to eliminate the uncertainty of behaviors. Similar to the Signal-to-Interference-plus-Noise Ratio (\emph{SINR}) in communication, this uncertainty can be measured as follows:
\begin{align}
SINR(b, P_I)=\frac{\sum_{p_i \in P_{I}}Sim(b,p_i)^{-\alpha}\cdot \exp(-H_{p_i})}{N_0+\sum_{p_j \in \mathcal{P}-P_{I}}Sim(b,p_j)^{-\alpha}\cdot \exp(-H_{p_j})},\nonumber
\end{align}
where $P_{I}$ denotes the set of behavior patterns of the user to be verified, $\mathcal{P}$ represents the set of all behavior patterns, and $H_{p_i}$ represents the behavior entropy of the current matched user. The function $Sim()$ measures the similarity between behaviors. The impact on behavioral authentication performance arises from interference by users with similar behavior patterns and the noise inherent in the data itself. For different levels of behavioral authentication, the elements contained in $P_{I}$   have distinct meanings.
%where $P_{I}$ 代表待认证用户行为模式的集合，$P$代表所有行为模式的集合，$H_{p_i}$ 代表当前匹配用户的行为熵。$Sim()$ 度量了行为之间的相似性。对于不同级别的行为认证而言，$P_{I}$和$P$ 包含的元素具体含义是不同的。
Taking behavioral identity authentication as an example, the set  $P_{I}$  degenerates into a singleton set, containing only the user with a deterministic behavior pattern.  The signal strength of behavioral identity authentication can be measured by calculating the stability and similarity between the given behavior and the behavior pattern of the user to be verified.
% as well as the  between the given behavior and the behavior pattern of the user to be verified.

Building upon the formalization for behavioral authentication described above, we conduct a preliminary exploration of the fundamental limits of data utility for behavioral authentication \cite{yang2020fundamental}. Specifically, from the perspective of data distribution, we introduce a data utility function based on conditional entropy. We analogize behavioral authentication to the problem of signal transmission in communication channels. By establishing the relationship between conditional entropy and authentication accuracy, we derive upper bounds on accuracy for behavioral authentication: the Shannon upper bound and R{\'e}nyi upper bound. The former is based on Shannon entropy and is obtained by transforming the Fano inequality; the latter utilizes conditional entropy defined on R{\'e}nyi entropy and is derived by applying Jensen's inequality and the principle of maximum discrete entropy. When the order of R{\'e}nyi entropy is greater than 1, the R{\'e}nyi upper bound can be expressed as follows:
%\begin{equation}
%	\begin{split}
%		Accuracy \leq 1 - \frac{H_{\alpha}(Y|\bm{X})- H_S(e)}{\log_2(N-1)},\\
%	\end{split}\nonumber
%\end{equation}	
    where $\bm{X}$ denotes the attribute combination, $ N$ refers to the number of categories of the label $ Y $, $ e $ denotes the average authentication error probability, $ H_{\alpha} $ denotes the R{\'e}nyi entropy, and $ H_S$ denotes the Shannon entropy. When the order of R{\'e}nyi entropy tends to $1$, the R{\'e}nyi upper bound degenerates to the Shannon upper bound.
On the real-world business dataset after privacy protection, we obtain the achievable bounds on accuracy for behavioral authentication using state-of-the-art and representative ensemble learning and deep learning models. By comparing the theoretical upper bound with the achievable bound, we observe that the theoretical upper bound is higher than the achievable bound, and they are very close to each other. This indicates that the theoretical upper bound provides valuable insights for optimizing the achievable bound.

Furthermore, in the field of communication, there have been initial advancements in semantic communication paradigms \cite{dong2022semantic, dai2022nonlinear}, which improve the transmission efficiency and reduce the latency  of communication systems. In the future, incorporating semantic information into our proposed formalization method for behavioral authentication holds the potential to push the fundamental limits of behavioral authentication.

\section{Discussions and Conclusion}\label{conclusion}
%\vspace{-0.1in}
%-----------------------------------------
\subsection{Applications}
Behavioral authentication has already been applied in practical applications such as financial risk control, healthcare, and intelligent transportation.

In financial transaction services, behavioral authentication can deduce invariant behavioral patterns from changing behavioral data. By integrating multiple factors such as the financial chain, consumption patterns, timing, and location of ordinary consumers, fragmented information is used to accomplish behavioral modeling of consumers. This forms the basis for constructing a comprehensive risk prevention and control system. Behavioral authentication technologies have been  implemented in the risk prevention systems of banks and Ant Group\cite{roa2021super,wang2022wrongdoing}.  It enables early detection and prevention of fraudulent activities.

In  healthcare applications, smart wearable devices from companies like Apple and Huawei use your behavioral data to assess your physical health status \cite{kunchay2020watchover,xiao2024fatigue}. Behavioral authentication provides continuous monitoring and important guidance for the user's health.  The industry has witnessed the emergence of behavioral capture and analysis products, such as fall detection vests for the elderly. When the walking patterns, accelerations, and other behavioral features of elderly individuals deviate from their normal patterns, the vest automatically activates a protective mode to mitigate the impact of a potential fall and prevent injuries \cite{ramachandran2020survey,pei2021elderly}.
%The implementation of behavioral authentication offers peace of mind to caregivers and loved ones, as it enhances the safety and well-being of the elderly.

In  intelligent transportation services, leading automotive companies such as Byd  and Tesla have implemented real-time behavior detection of drivers in their onboard systems \cite{xun2021deep,ma2023exploring}.
%The onboard system intervenes when hazardous lane changes, overtaking at an unsafe distance, and other  non-benign behaviors are detected.
 When non-benign behaviors are detected, such as abrupt lane changes without signaling or tailgating at an unsafe distance, the onboard system intervenes to mitigate potential risks.  This intervention can take the form of audible warnings, visual alerts, or even automated corrective actions, such as gentle steering corrections or adaptive cruise control adjustments.
The implementation of behavioral authentication in intelligent transportation services helps to cultivate safer driving habits and reduce the likelihood of accidents.

\subsection{Conclusion}
Technological advancements have ushered humanity into an unprecedented era of artificial intelligence.
Behavioral authentication has emerged as a promising authentication solution in many scenarios and been successfully and widely applied in practice. It has gradually become a fundamental problem in the field of authentication.
This work summarizes the background and applications of behavioral authentication.
In particular, it introduces concept of behavioral authentication including behavioral identity authentication, behavioral conformity authentication, and behavioral benignity authentication. The paper presents a comprehensive review of work conducted in these three levels of behavioral authentication, establishes a clear framework for categorization, and summarizes their corresponding characteristics and issues. The main challenges in current behavioral authentication are analyzed, and key research directions for the future are pointed out including fine-grained behavior modeling, distributed behavioral authentication and formalization for behavioral authentication.
Obviously, with the increasing sophistication of artificial intelligence in various fields and  a greater emphasis on user experience, behavioral authentication will display even stronger vitality and play an increasingly important role.

%\section*{Acknowledgement}
%This work was supported in part by the National Key Research and Development Program of China under Grant YS2022YFB4500205 and in part by the Shanghai Science and Technology Innovation Action Plan Project under Grant 22511100700.

%This work was supported in part by the National Key Research and Development Program of China under Grant 2022YFB4501704 and in part by the Shanghai Science and Technology Innovation Action Plan Project under Grant 22511100700.
%%
\conflict

The author declare no conflict of interest.
\dataavailability
 No data are associated with this article.
\authorcontrib
%
%This is Authors Contributions text here. If only one author as this article, delete this part. For more than one author articles, please have contribution statement for each author.
Cheng  Wang and Changjun Jiang designed the research. Cheng  Wang and Hao Tang articulated the concept of behavioral authentication and divided it into three different levels, and wrote the paper.
Hangyu Zhu established a preliminary framework for fine-grained behavioral modeling in the future work.
Junhan Zheng participated in the literature review of existing studies.
All authors read and approved the final manuscript.
\acknowtext
We thank the anonymous reviewers for their helpful comments.

%This is Acknowledgements text here.
%王成and 蒋昌俊 designed the research. 王成和唐昊阐述了行为认证的理念并把他分为三种不同级别的认证，撰写了全文。
%朱航宇建立了未来工作部分的一体化行为建模的初步方案。
%郑俊韩参与了现有文献的调研与梳理

%
\funding
This works was supported in part by the National Natural Science Foundation of China (NSFC) under Grant 62372328, in part by the Program of Shanghai Academic Research Leader under Grant 22XD1423700, in part by the National Key Research and Development Program of China under Grant 2022YFB4501704, in part by the Shanghai Science and Technology Innovation Action Plan Project under Grant 22511100700, in part by the Leadership Project under the Oriental Talent Program, and in part by the Open Fund of Key Laboratory of Industrial Internet of Things and Networked Control, Ministry of Education, under Grant 2021FF08.
\bibliographystyle{sands-r}
\bibliography{ref_miner}

\vitanew{\vita{3}{l}[h]{\includegraphics[width=20mm]{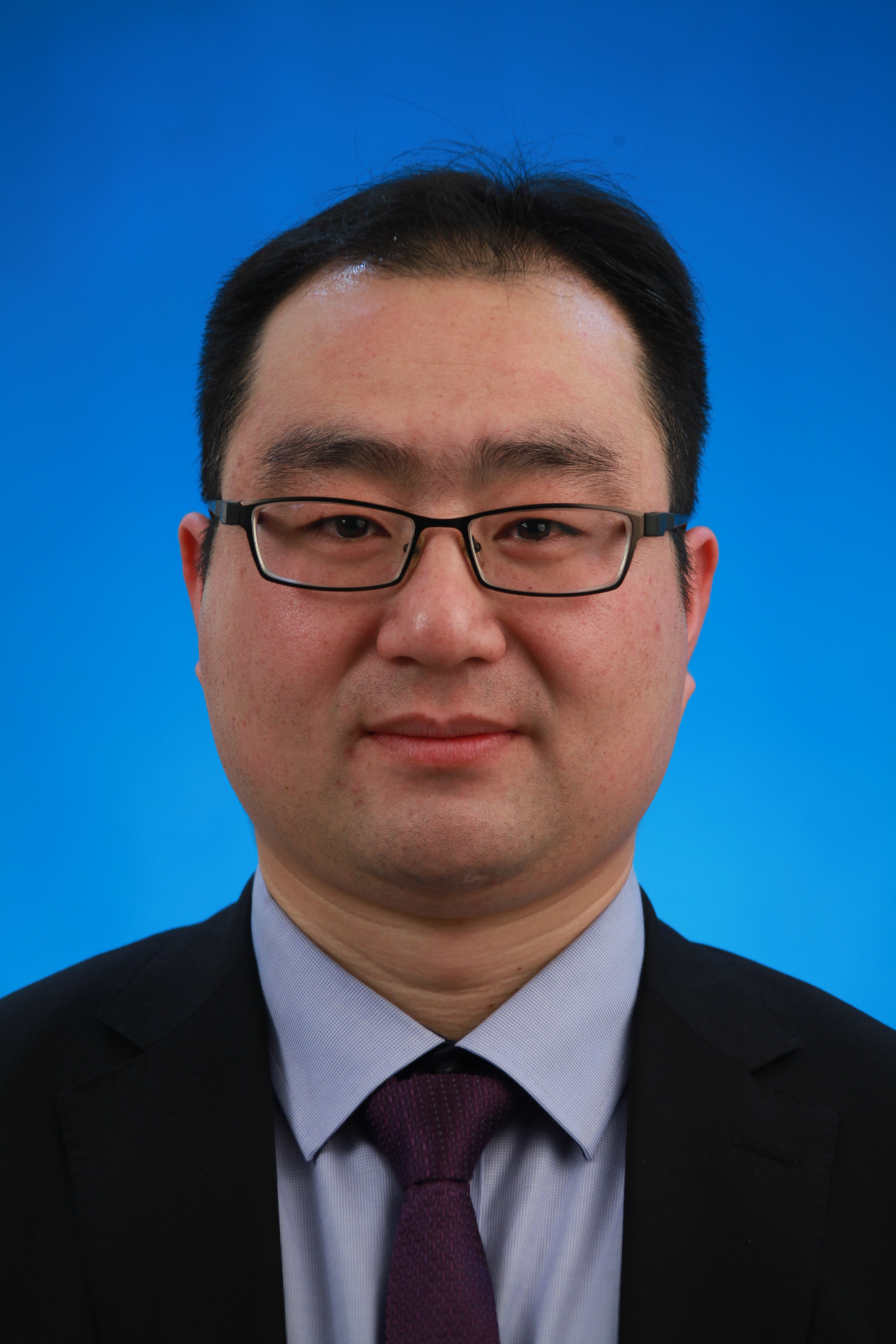}}
{Cheng Wang}{received the master's degree at Department of Applied Mathematics from Tongji University, Shanghai, China, in 2006 and the Ph.D. degree in Department of Computer Science at Tongji University in 2011. He is currently a Professor and the
Head of the Department of Computer Science at Tongji University. His research interests include
distributed learning, cyberspace security and intelligent information services.}}

\vitanew{\vita{3}{l}[h]{\includegraphics[width=20mm]{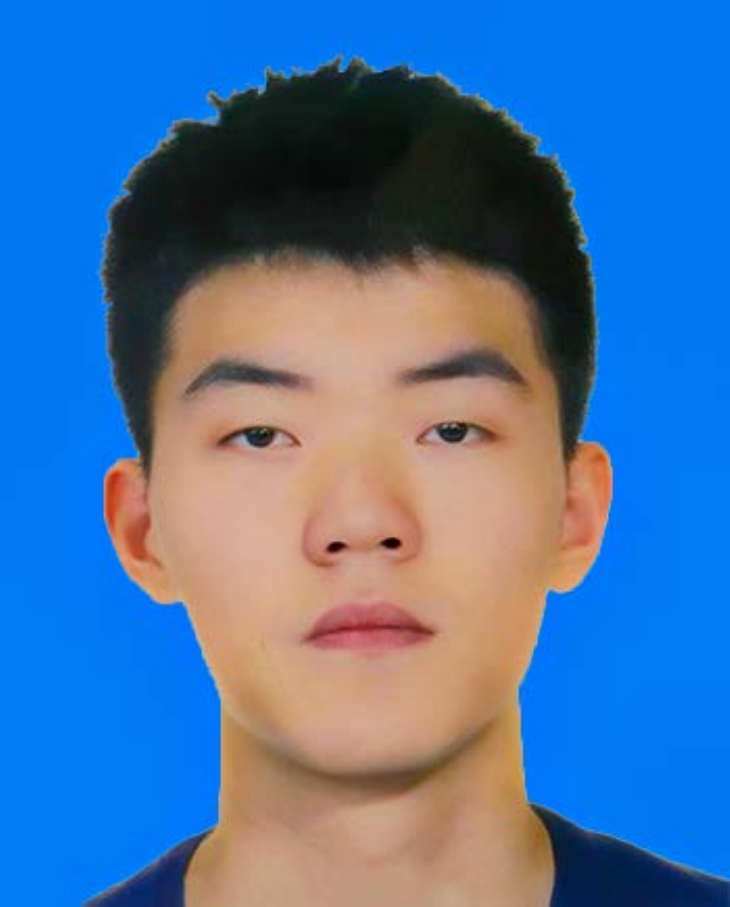}}
{Hao Tang}{received his bachelor degree of engineering at Department of Computer Science from Chongqing University of Posts and Telecommunications in 2020. He is currently pursuing the Ph.D. degree at Department of Computer Science at Tongji University in Shanghai, China. His research interests include privacy-preserving learning and  intelligent information services.}}

\vitanew{\vita{3}{l}[h]{\includegraphics[width=20mm]{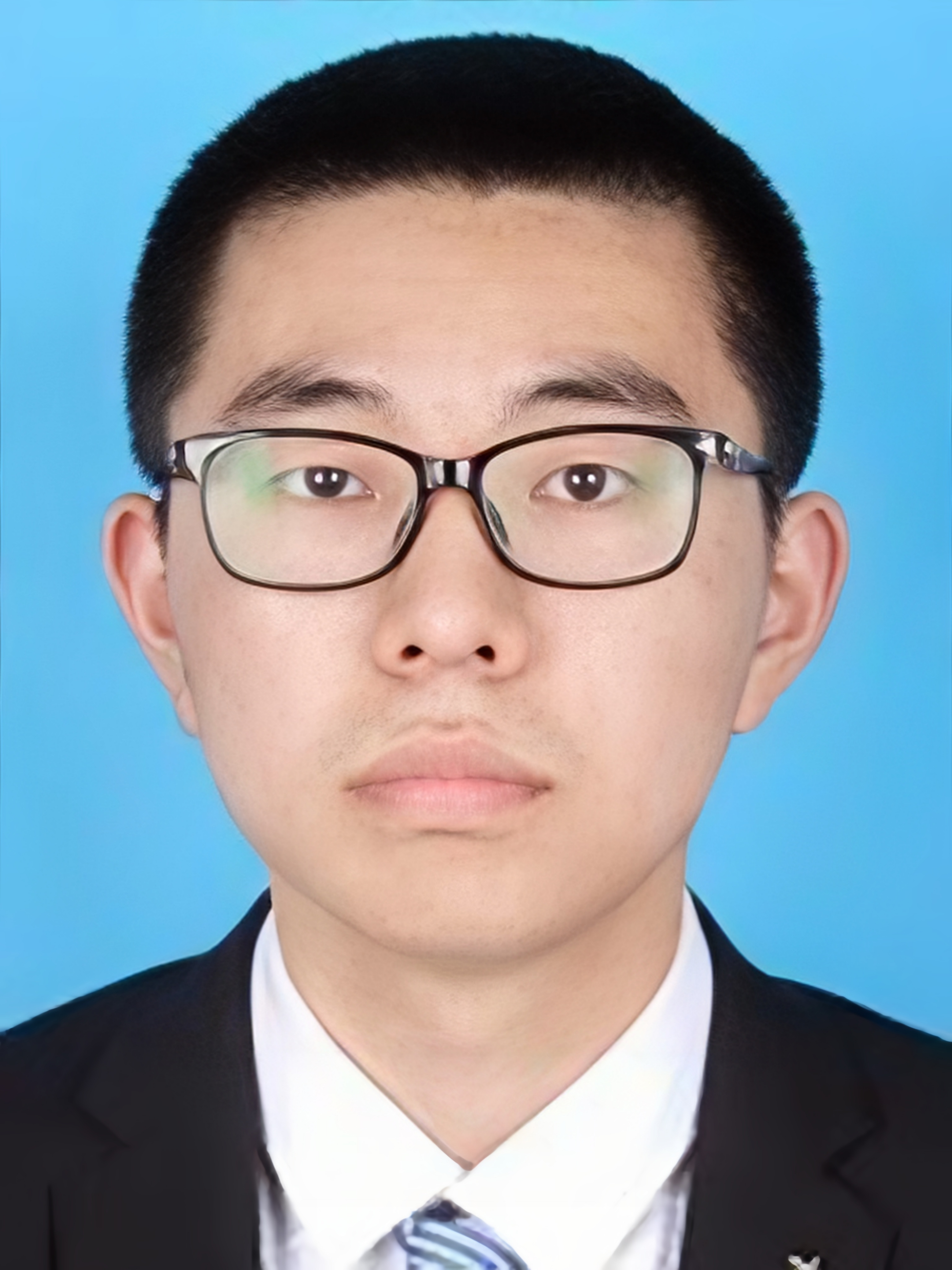}}{Hangyu Zhu}{received the master's degree from the
Department of Computer Science and Technology,
Tongji University, Shanghai, China, in 2021,
where he is currently pursuing the Ph.D. degree with
the Department of Computer Science. His research
interests include anomaly detection and network
representation learning.}}

\vitanew{\vita{4}{l}[h]{\includegraphics[width=20mm]{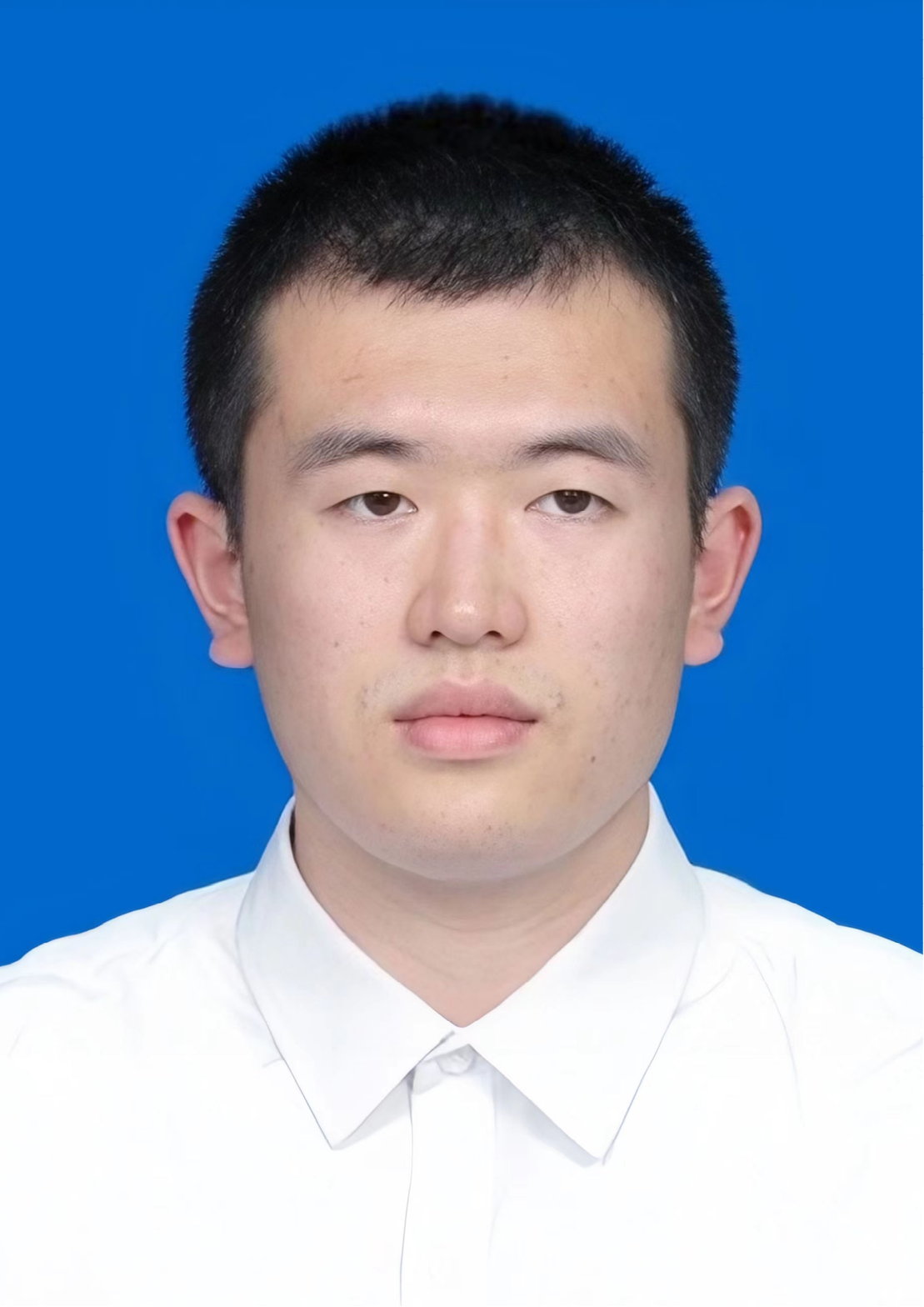}}{Junhan Zheng} {received his  bachelor degree in computer science from Tongji University, Shanghai, China, in 2023, where he is currently working towards the master's degree from the master's degree with the Department of Computer Science and Technology. His research interests include anomaly detection and attack investigation.}}
\vspace{-0.05in}

\vitanew{\vita{4}{l}[h]{\includegraphics[width=20mm]{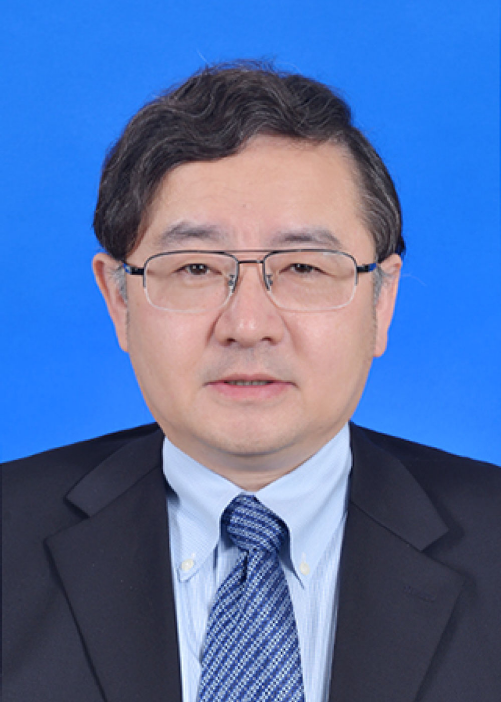}}{Changjun Jiang} {received the Ph.D. degree from the Institute of Automation, Chinese Academy of Sciences, Beijing, China, in 1995.
 He is currently the Leader of the Key Laboratory of the Ministry of Education for Embedded System and Service Computing, Tongji University, Shanghai, China. He is also an Honorary Professor with Brunel University London, Uxbridge, England.
 Dr. Jiang is also an IET Fellow. He is an Academician of Chinese Academy of Engineering. His research interests include concurrence theory, Petri nets, formal verification of software, cluster, grid technology, intelligent transportation systems, and service-oriented computing. }}
\end{document}